\documentclass[9pt,twoside]{article}
\usepackage{amssymb}
\usepackage{latexsym}
\usepackage{epic}
\usepackage{bez123}
\usepackage{graphicx}
\usepackage{ifpdf}
\makeatletter

\@addtoreset{equation}{section} \makeatother
\newtheorem{theorem}{Theorem}[section]
\newtheorem{convention}[theorem]{Convention}
\newtheorem{remark}[theorem]{Remark}
\newtheorem{lemma}[theorem]{Lemma}
\newtheorem{proposition}[theorem]{Proposition}
\newtheorem{example}[theorem]{Example}
\newtheorem{corollary}[theorem]{Corollary}
\newtheorem{definition}[theorem]{Definition}

\newcommand{\relst}{\sim_{st}}
\newcommand{\cvd}{\ \rule{0.5em}{0.5em}}
\newcommand{\be}{\begin{equation}}
\newcommand{\ee}{\end{equation}}

\newcommand{\N}{{\mathbb N}}

\newcommand{\R}{{\mathbb R}}

\newcommand{\ben}{\begin{enumerate}}
\newcommand{\een}{\end{enumerate}}
\newcommand{\bit}{\begin{itemize}}
\newcommand{\eit}{\end{itemize}}
\newcommand{\edoc}{\end{document}}

\newcommand{\aux}{{\rm aux}}
\newcommand{\talpha}{\tilde{\alpha}}

\newcommand{\ca}{op}
\newcommand{\cc}{cl}

\newcommand{\bstrain}{ST}
\newcommand{\bstrainp}{ST^+}
\newcommand{\bstrainm}{ST^-}

\newcommand{\replace}[2]{#2}

\newcommand{\bpcc}[1]{b^+_{{#1},\cc}}
\newcommand{\bpca}[1]{b^+_{{#1},\ca}}
\newcommand{\bmcc}[1]{b^-_{{#1},\cc}}

\newcommand{\apf}{\hat{{\cal J}}}
\newcommand{\app}{\check{{\cal J}}}
\newcommand{\ap}{{\cal J}}
\newcommand{\apt}{\overline{j}}
\newcommand{\aptq}{{\cal J}}

\newcommand{\Lin}{{\cal L}}

\newcommand{\LI}{{\rm LI}}
\newcommand{\LS}{{\rm LS}}

\newcommand{\LQG}{L_{\sim}}
\newcommand{\LQ}{\hat{L}_{\sim_{st}}}
\newcommand{\LQQ}{L_{\sim_{st}}}

\newcommand{\columna}{column}
\newcommand{\raspa}{skeleton}
\newcommand{\espina}{bone}

\begin{document}
\parindent=5mm
\date{}

\medskip


\title{{\bf\LARGE  Computability of the causal boundary by using isocausality}}

\author{{\bf\large J.L. Flores$^*$,
J. Herrera$^{*\dagger}$,
M. S\'anchez$^\dagger$}\\
{\it\small $^*$Departamento de \'Algebra, Geometr\'{\i}a y Topolog\'{\i}a,}\\
{\it \small Facultad de Ciencias, Universidad de M\'alaga,}\\
{\it\small Campus Teatinos, 29071 M\'alaga, Spain}\\
{\it\small $^\dagger$Departamento de Geometr\'{\i}a y Topolog\'{\i}a,}\\
{\it\small Facultad de Ciencias, Universidad de Granada,}\\
{\it\small Avenida Fuentenueva s/n, 18071 Granada, Spain}}

\maketitle

\begin{abstract}
Recently, a new viewpoint on the classical c-boundary in
Mathematical Relativity has been developed, the relations of this
boundary with the conformal one and other classical boundaries
have been analyzed, and its computation in some  classes of
spacetimes, as the standard stationary ones, has been carried out.

In the present paper, we consider the notion of {\em isocausality}
given by Garc\'{\i}a-Parrado and Senovilla, and introduce a
framework to carry out isocausal comparisons with standard
stationary spacetimes. As a consequence, the qualitative behavior
of the c-boundary (at the three levels: point set, chronology and
topology) of a wide class of spacetimes, is obtained.
\end{abstract}
\begin{quote}
\end{quote}
\begin{quote}


\end{quote}

\newpage
\tableofcontents
\newpage

\section{Introduction}

Recently, a new viewpoint and definition of the notion of {\em
causal boundary} (c-boundary, for short) $\partial V$ of a
spacetime $V$ have been developed in \cite{FHSconf}. In the case
of spacetimes which are (conformal to) standard stationary ones,
such a boundary has been extensively studied in \cite{FHSst},
where the c-boundary is also carefully connected with other
boundaries which appear naturally in Differential Geometry.

The machinery introduced in \cite{FHSconf, FHSst} is enough to
compute explicitly the c-boundary in many cases. Nevertheless,
from a practical viewpoint, the c-boundary points may present
quite a few of subtle (and bothering) possibilities for general
spacetimes. This also happens in other simple boundaries. For
example, the elementary Cauchy boundary of a (positive-definite)
Riemannian manifold, regarded as a metric space, may be
non-locally compact
---and such a property can be transmitted to the c-boundary. Of
course, one does not expect that, in realistic spacetimes, all
type of pathological mathematical properties will hold. In
principle, a qualitative description of the c-boundary (with some
criteria to understand at what extent some types of pathologies
may exist) would be enough for many purposes. In the present
article, this idea will be carried out by comparing the c-boundary
of broad classes of spacetimes with the c-boundary of well-behaved
standard stationary spacetimes. For this aim, the notion of {\em
isocausality} in the sense introduced by Garc\'{\i}a-Parrado and
Senovilla \cite{GpScqg03} (see also \cite{GP-Sa,GSa,GSb,GpScqg05})
will be used sistematically.

Roughly speaking, two spacetimes are called isocausal when, by
using a pair of diffeomorphisms, the timecones of the first one
can be seen inside the timecones of the second one, and viceversa
(see Definition \ref{defiso}). This includes the case of conformal
equivalence (where a single diffeomorphism and its inverse are
used) but isocausality is much more flexible. Even though this
notion is simple, isocausality is also a subtle concept. For
example, most of the steps of the so-called {\em causal ladder of
spacetimes} are preserved by isocausal spacetimes \cite{GpScqg03} but not
all of them \cite[Section 3.2]{GP-Sa}. And, recently the
authors exhibited an example of two isocausal spacetimes with
different c-boundaries \cite{FHSnota}.

In spite of this last result (or perhaps because of it), we will
show that the notion of isocausality is useful in order to analyze
the c-boundary of some general classes of spacetimes. In fact,
when a spacetime $V$ is isocausal to a simple standard stationary
one $V_{\cc}$ (whose c-boundary $\partial V_{\cc}$ is known from
\cite{FHSst}), a very interesting information on the c-boundary
$\partial V$ of the original spacetime $V$ is obtained. Roughly
speaking, $\partial V$ will be at least as rich as $\partial
V_{\cc}$. This means that $\partial V$ contains the elements of
$\partial V_{\cc}$ but perhaps some of these elements are
``enlarged'' ---more precisely,  $\partial V_{\cc}$ is isomorphic
to certain quotient of $\partial V$. Under simple hypotheses
(expected to hold in the physically relevant cases) such
enlargements are reasonably controlled.

This paper is organized as follows. In Section \ref{s2} we recall
the notion of c-boundary introduced in \cite{FHSconf}, as well as
some general elements of Finsler manifolds which are necessary for
the computation of the c-boundary 
of a stationary spacetime. In Section \ref{s3} we consider the
expression of a standard stationary spacetime endowed with an
additional time dependent factor $\tilde\alpha(t)$. This
expression may be  interesting by itself (for example, in order to
write the c-boundary  when the condition (\ref{intcond}) does not
hold, extending the results in \cite{FHSst}). In this paper, such
an expression will allow to make isocausal comparisons easily. For
this purpose,
we rewrite the  known expressions of the c-boundary of a standard
stationary spacetime (computed explicitly in \cite{FHSst} for the
case $\tilde \alpha(t) \equiv 1$) taking into account such a
factor. In Section \ref{s4}, after explaining the notion of
isocausality, the framework for isocausal comparison with a
standard stationary spacetime is introduced. Choosing $\tilde
\alpha \equiv 1$ and $\tilde \alpha \equiv \alpha <1$ we obtain
two metrics $g_{\cc}, g_{\ca}$ which satisfy: (i) $g_{\cc} \prec_0
g_{\ca}$, i.e., the timecones of the former are narrower than
those of the latter, and (ii) both metrics are conformal and,
thus, isocausal. Any metric $g$ with $g_{\cc} \prec_0 g \prec_0
g_{\ca}$ is then isocausal to $g_{\cc}$ (and $g_{\ca}$) and, so,
our  aim will be to relate the c-completions $\bar V_{\cc}$ and
$\bar V$ for $g_{\cc}$ and $g$, resp. According to \cite{FHSst},
as $g_{\cc}$ is standard stationary, its c-boundary
$\partial_{\cc} V$ can be constructed by using a Finsler metric
$F$ (the {\em Fermat metric}), which will be also compared in this
section with a parameterized Finsler metric $F_t$ associated to
$g$. In Section \ref{s5} the relation between the future
boundaries $\hat\partial_{\cc} V$ and $\hat
\partial V$ (as well as between the past boundaries
$\check\partial_{\cc} V$ and $\check
\partial V$) is analyzed, and in Section \ref{s6} the relation
between the full c-boundaries $\partial_{\cc} V$ and $\partial V$
is achieved. In the remainder of this Introduction, we give a
heuristic explanation of these relations.

\begin{figure}
\centering
\ifpdf
  \setlength{\unitlength}{1bp}%
  \begin{picture}(171.64, 285.62)(0,0)
  \put(0,0){\includegraphics{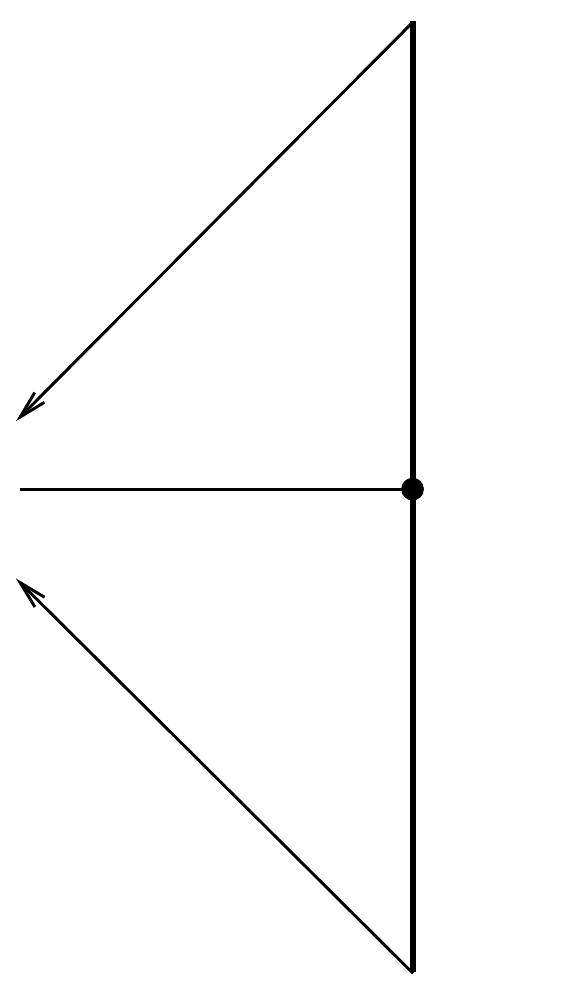}}
  \put(95.65,148.44){\fontsize{14.23}{17.07}\selectfont $P_0$}
  \put(7.22,137.85){\fontsize{8.54}{10.24}\selectfont $x$}
  \put(121.46,268.86){\fontsize{8.54}{10.24}\selectfont $t$}
  \end{picture}%
\else
  \setlength{\unitlength}{1bp}%
  \begin{picture}(171.64, 285.62)(0,0)
  \put(0,0){\includegraphics{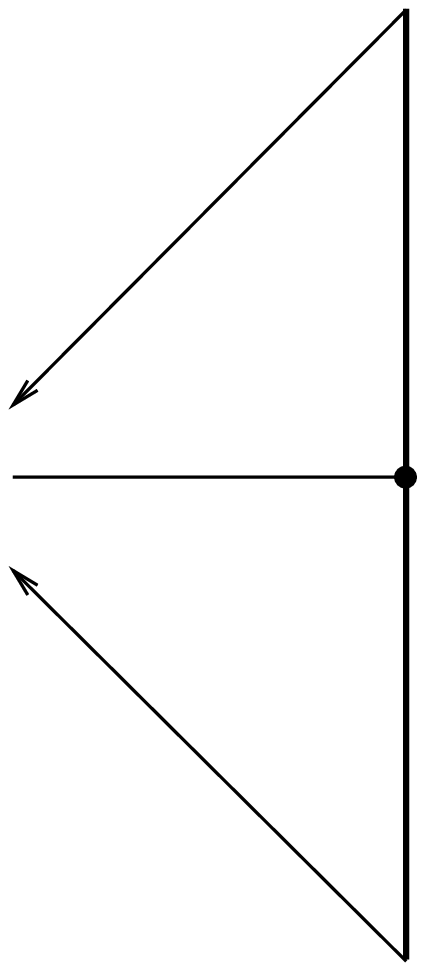}}
  \put(95.65,148.44){\fontsize{14.23}{17.07}\selectfont $P_0$}
  \put(153.22,137.85){\fontsize{8.54}{10.24}\selectfont $x$}
  \put(121.46,268.86){\fontsize{8.54}{10.24}\selectfont $t$}
  \end{picture}%
\fi \caption{\label{fig1} {\em The c-boundary of
$(\R\times\R^-,-dt^2+dx^2)$.}\newline The c-boundary is composed
by a timelike line (the vertical segment) and two lightlike lines
(which will not be relevant here).}
\end{figure}

\begin{figure}
\centering
\ifpdf
  \setlength{\unitlength}{1bp}%
  \begin{picture}(316.40, 195.23)(0,0)
  \put(0,0){\includegraphics{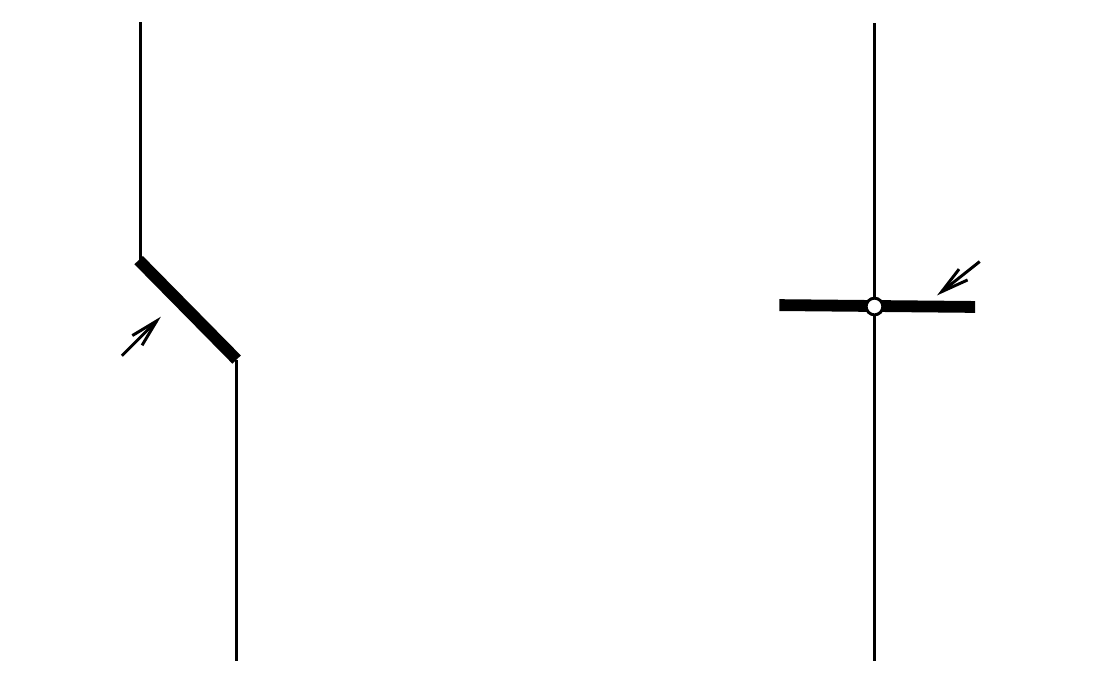}}
  \put(109.01,105.49){\fontsize{14.23}{17.07}\selectfont $\Rightarrow$}
  \put(5.67,84.41){\fontsize{8.54}{10.24}\selectfont $St(P_0)$}
  \put(276.23,122.27){\fontsize{8.54}{10.24}\selectfont $St(P_0)$}
  \end{picture}%
\else
  \setlength{\unitlength}{1bp}%
  \begin{picture}(316.40, 195.23)(0,0)
  \put(0,0){\includegraphics{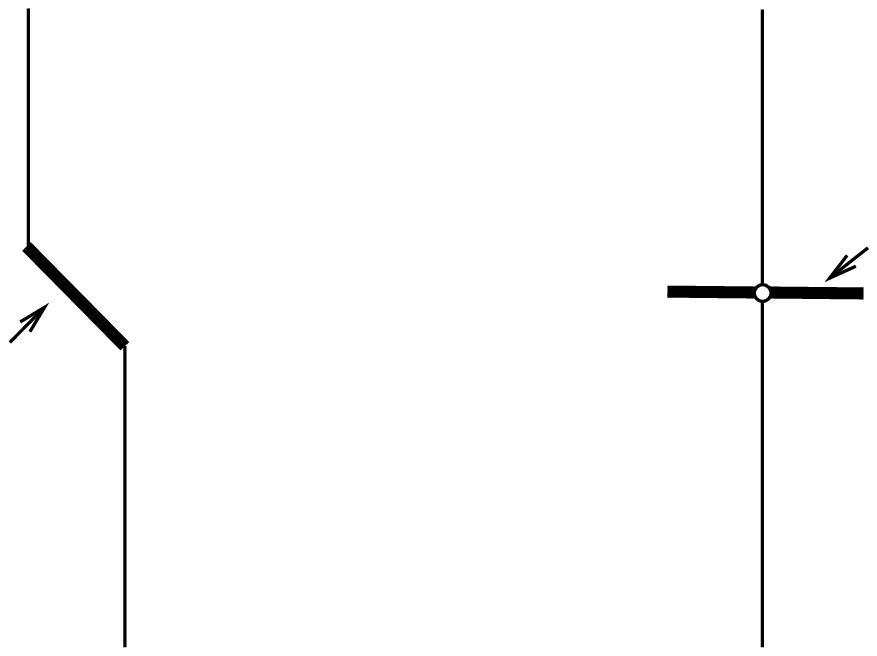}}
  \put(109.01,105.49){\fontsize{14.23}{17.07}\selectfont $\Rightarrow$}
  \put(5.67,84.41){\fontsize{8.54}{10.24}\selectfont $St(P_0)$}
  \put(276.23,122.27){\fontsize{8.54}{10.24}\selectfont $St(P_0)$}
  \end{picture}%
\fi \caption{\label{fig2}{\em C-boundary of $(V,g)$.}\newline The
strain St$(P_0)$ is drawn on the left as in \cite{FHSnota} and on
the right, more conceptually, as an horizontal segment (a
``bone'').}
\end{figure}

\begin{figure}
\centering
\ifpdf
  \setlength{\unitlength}{1bp}%
  \begin{picture}(117.09, 235.34)(0,0)
  \put(0,0){\includegraphics{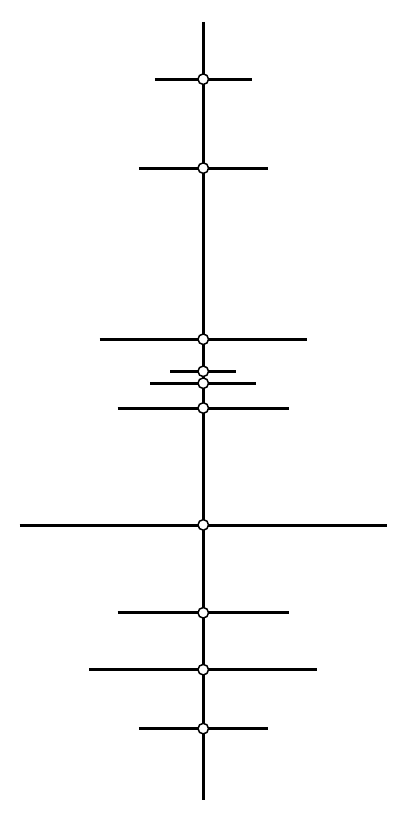}}
  \end{picture}%
\else
  \setlength{\unitlength}{1bp}%
  \begin{picture}(117.09, 235.34)(0,0)
  \put(0,0){\includegraphics{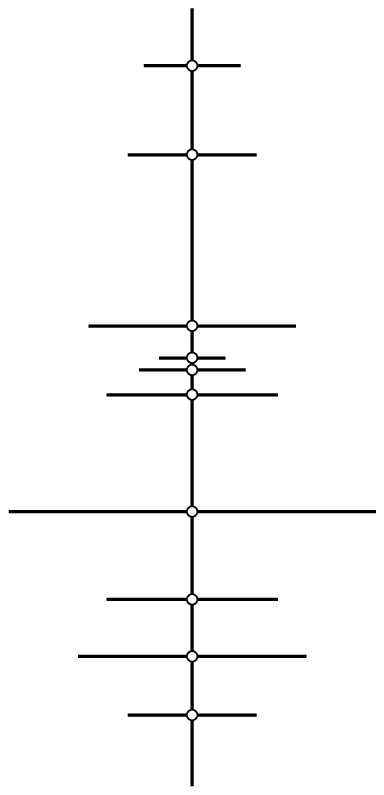}}
  \end{picture}%
\fi \caption{\label{fig3}{{\it Conceptual view of the ``{\raspa}''
structure of the c-boundary.}\newline The vertical
segment (``{\columna}'') represents the quotient space obtained by collapsing each strain (drawn as a horizontal one, a ``{\espina}''). This conceptual view of the strains as bones does not represent any topological nor causal structure as, in general, they could present different (and non-equivalent) structures (see Figure \ref{figstraingeneral}).}}
\end{figure}

The possibility that two isocausal spacetimes may have different
c-boundaries is explained carefully in  \cite{FHSnota}. There, we
considered three metrics $g_{\cc}\prec_0 g \prec_0 g_{\ca}$ on
$\R\times \R^-$ such that $g_{\cc}$ and $ g_{\ca}$ are conformal
to the usual one $g_0= -dt^2+dx^2$. So, the c-boundary
$\partial_{\cc}V (\equiv \partial_{\ca} V)$ is homeomorphic and
naturally identifiable to the conformal boundary of $(\R\times
\R^-, g_0)$ in $\R^2$, i.e. the line $x=0$, plus two lightlike
lines that will be irrelevant for our discussion  (Figure
\ref{fig1}).
The metric $g$ and its boundary $\partial V$ requires a somewhat
involved construction (see the details in \cite{FHSnota}). When
one considers the boundary point $P_0=(0,0)\in
\partial_{\cc}V$, and try to relate it with some point of  $\partial V$, 
one
realizes that $P_0$ is stretched in a bigger piece of boundary,
which we called the {\em strain} of $P_0$, Str$(P_0) (\subset
\partial V)$. In this particular
example, Str$(P_0)$ is naturally a lightlike segment (see
Figure \ref{fig2} left) but, in general, its structure may be more
complicated. So, we prefer to represent it conceptually as in
Figure \ref{fig2} right.
Obviously, one could refine this example, introducing more strains
for other points, obtaining so the ``{\raspa}'' structure of
Figure \ref{fig3}. Clearly,  $\partial V_{\cc}$ is recovered by
collapsing each strain of $\partial V$ to a single point.

In Sections \ref{s5}, \ref{s6} we show at what extent this simple
picture of $\partial V$ and $\partial_{\cc}V$ holds in general.
With this aim, we impose two type of hypothesis:
\begin{itemize}\item Hypotheses to ensure
that the metric $g$ is reasonably close to $g_{\cc}$ and $g_{\ca}$
for large $|t|$: the integral conditions (\ref{e4}) and
(\ref{e4'}) on the  factor $\alpha$ mentioned above (the former
condition  for the future boundary and the latter for the past
one). These conditions are satisfied if, for example, $\alpha$
behaves as $1+O(1/|t|^{1+\epsilon})$ for large $|t|$ and some
$\epsilon>0$. \item
Hypotheses to ensure that the model boundary $\partial_{\cc}V$ is
not too pathological. Concretely, the canonic generalized distance
$d^+$ of the Finsler metric associated to $g_{\cc}$ will satisfy:
(a) its Cauchy completion $M_C^+$ is locally compact, and (b) the
extension $d_Q^+$ of $d^+$ to the Cauchy boundary is still a
generalized distance. For example, if we consider as the metric
$g_{\cc}$ a static one ($\omega \equiv 0$ in formula
(\ref{exp1})), the Finsler metric becomes Riemannian, and the
generalized distance $d^+$ becomes  a usual (symmetric) distance.
So, condition (b) is automatically satisfied. On the other hand,
condition (a) just prevents  an undesirable property even for a
Riemannian manifold (see the Riemannian example 4.9  and the
Finslerian examples explained in Remark 3.26, both in reference
\cite{FHSst}, for some pathological samples where hypotheses (a)
or (b) are not satisfied).
\end{itemize}
Under these hypotheses, in Section \ref{s5} we show that the
partial boundaries $\hat \partial V, \check \partial V$ behave
essentially as in the heuristic example above, that is,
$\hat{\partial} V$ must contains $\hat{\partial}_{\cc} V$ in the
sense that a quotient $\hat{\partial} V/\sim_{st}$ is identifiable
to $\hat{\partial}_{\cc} V$ (this occurs at two levels, as point
set and topologically, recall Theorems \ref{apfc}, \ref{apfc'} and
Remark \ref{rf1}).
Finally, in Section \ref{s6} we show that these considerations can
be extended to the (total) c-boundary (see Theorem \ref{theoremf},
plus Remark \ref{rf2} for some subtleties at the point set level).
Moreover, this includes also considerations about the causal and
chronological relations in the completions $\overline{V},
\overline{V}_{\cc}$ which are related to the quotient space
$\overline{V}/\sim$ (see Theorem \ref{f} and Definition
\ref{defcauqu}).


\section{Preliminaries}
\label{s2}

Along this paper we will use typical background and terminology in
Lorentzian Geometry as in \cite{BEE, MS, O}. A {\em spacetime} $V$
will be a time-oriented connected smooth Lorentzian $(-,+, \dots , +)$
manifold, also denoted $(V,g)$ abusing of notation.
 A tangent
vector $v\in T_{p}V$, $p\in V$ is called {\em timelike} (resp.
{\em lightlike}; {\em causal}) if $g(v,v)<0$ (resp. $g(v,v)=0$,
$v\neq 0$; $v$ is either timelike or lightlike). A causal vector
is called {\em future} or {\em past-directed} if it belongs to the
future or past cone. Accordingly, a smooth curve
$\gamma:I\rightarrow V$ ($I$ real interval) is called timelike,
lightlike, causal and future or
past-directed if so is $\dot{\gamma}(s)$ 
for all $s\in I$.

Two events $p,q\in V$ are {\em chronologically related} $p\ll q$
(resp. {\em causally related} $p\leqslant q$) if there exists some
future-directed timelike (resp. either future-directed causal or
constant) curve from $p$ to $q$.
If $p\neq q$ and $p\leqslant q$ but $p\not\ll q$, then $p$ is said {\em
horis\-mo\-ti\-cal\-ly related} to $q$. The {\em chronological
past} (resp. {\em future}) of $p$, $I^{-}(p)$ (resp. $I^{+}(p)$)
is defined as:
\[
I^{-}(p)=\{q\in V: q\ll p\}\qquad(\hbox{resp.}\;\; I^{+}(p)=\{q\in
V: p\ll q\}).
\]
The chronological past $I^-(\gamma)=\cup_{s\in I}I^-(\gamma(s))$
(resp. future $I^+(\gamma)=\cup_{s\in I}I^+(\gamma(s))$) of
future-directed (resp. past-directed) timelike curves
$\gamma:I\rightarrow V$ will play a relevant role.

Along the paper several (time-oriented) Lorentzian metrics on the same manifold will be used.
To avoid confusions, we will introduce a subindex to indicate the
corresponding metric we are considering; say, $\ll_{\tilde{g}}$ or
$I^{-}_{\tilde{g}}$ for the metric $\tilde{g}$ on $V$.

\subsection{C-boundary of spacetimes}\label{d}
The {\em c-completion} of spacetimes is constructed by adding {\em
ideal points} to the spacetime in such a way that any timelike
curve in the original spacetime acquires some endpoint in the new
space \cite{GKP}. To formalize this construction, which will be
conformally invariant and applicable to any {\em strongly causal}
spacetime, previously we need to introduce some basic notions.

A non-empty
subset $P\subset V$ is called a {\em past set} if it coincides
with its past; i.e. $P=I^{-}(P):=\{p\in V: p\ll q\;\hbox{for
some}\; q\in P\}$. The {\em common past} of $S\subset V$ is
defined by $\downarrow S:=I^{-}(\{p\in V:\;\; p\ll q\;\;\forall
q\in S\})$. In particular, the past and common past sets must be
open. A past set that cannot be written as the union of two proper
past sets
is called {\em indecomposable past} set, {\em IP}. An IP which
does coincide with the past of some point of the spacetime
$P=I^{-}(p)$, $p\in V$ is called {\em proper indecomposable past
set}, {\em PIP}. Otherwise, $P=I^{-}(\gamma)$ for some
inextendible future-directed timelike curve $\gamma$, and it is
called {\em terminal indecomposable past set}, {\em TIP}. The dual
notions of {\em future set}, {\em common future}, {\em IF}, {\em
TIF} and {\em PIF}, are defined just by interchanging the roles of
past and future in previous definitions.

To construct the {\em future} and {\em past c-completion}, first
we have to identify each {\em event} $p\in V$ with its PIP,
$I^{-}(p)$, and PIF, $I^{+}(p)$. This is possible in any {\em
distinguishing} spacetime, that is, a spacetime which satisfies
that two distinct events $p, q$ have distinct chronological
futures and pasts ($p\neq q \Rightarrow I^\pm (p) \neq I^\pm
(q)$). In order to obtain consistent topologies in the
c-completions, we will focus on a somewhat more restrictive class
of spacetimes, the {\em strongly causal ones}. These are
characterized by the fact that the PIPs and PIFs constitute a
sub-basis for the topology of the manifold $V$.

%

Now, every event $p\in V$ can be identified with its PIP,
$I^-(p)$. So, the {\em future c-boundary} $\hat{\partial}V$ of $V$
is defined as the set of all the TIPs in $V$, and  {\em the future
c-completion} $\hat{V}$ becomes the set of all the IPs:
\[
V\equiv \hbox{PIPs},\qquad \hat{\partial}V\equiv
\hbox{TIPs},\qquad\hat{V}\equiv \hbox{IPs}.
\]
Analogously, each $p\in V$ can be identified with its PIF,
$I^+(p)$. The {\em past c-boundary} $\check{\partial}V$ of $V$ is
defined as the set of all the TIFs in $V$, and  {\em the past
c-completion} $\check{V}$ is the set of all the IFs:
\[
V\equiv \hbox{PIFs},\qquad \check{\partial}V\equiv
\hbox{TIFs},\qquad\check{V}\equiv \hbox{IFs}.
\]

For the (total) c-boundary, the so-called S-relation comes into
play \cite{Sz}.
Denote $\hat{V}_{\emptyset}=\hat{V}\cup \{\emptyset\}$ (resp.
$\check{V}_{\emptyset}=\check{V}\cup \{\emptyset\}$). The
S-relation $\sim_S$ is defined in $\hat{V}_{\emptyset}\times
\check{V}_{\emptyset}$ as follows. First, in the case $(P,F)\in \hat{V}\times
\check{V}$, \be \label{eSz}  P\sim_S F \Longleftrightarrow \left\{
\begin{array}{l}
P \quad \hbox{is included and is a maximal IP into} \quad
\downarrow F
 \\
F \quad \hbox{is included and is a maximal IF into} \quad \uparrow
P.
\end{array} \right.
\end{equation}
By {\em maximal} we mean that no other $P'\in\hat{V}$ (resp.
$F'\in \check{V}$) satisfies the stated property and includes
strictly $P$ (resp. $F$). Recall that, as proved by Szabados
\cite{Sz}, $I^-(p) \sim_S I^+(p)$ for all $p\in V$, and these are
the unique S-relations (according to our definition (\ref{eSz}))
involving proper indecomposable sets. Now, in the case $(P,F)\in
\hat{V}_{\emptyset}\times \check{V}_{\emptyset}\setminus
\{(\emptyset,\emptyset)\}$, we also put \be \label{eSz2} P\sim_S
\emptyset, \quad \quad (\hbox{resp.} \; \emptyset \sim_S F )\ee if
$P$ (resp. $F$) is a (non-empty, necessarily terminal)
indecomposable past (resp. future) set that  is not S-related by
(\ref{eSz}) to any other indecomposable set; notice that
$\emptyset$ is never S-related to itself. Now, we can introduce
the notion of c-completion, according to \cite{FHSconf}:
\begin{definition}\label{d1}
As a point set, the {\em c-completion} $\overline{V}$ of a strongly causal spacetime $V$ is formed by all
the pairs $(P,F)\in
\hat{V}_{\emptyset}\times\check{V}_{\emptyset}$ such that
$P\sim_{S} F$. The {\em c-boundary} $\partial V$ is defined as
$\partial V:=\overline{V}\setminus V$, where $V\equiv
\{(I^{-}(p),I^{+}(p)): p\in V\}$.
\end{definition}

%


The chronological relation $\ll$ of the spacetime is extended to
the c-completion in the following way. We say that $(P,F),
(P',F')\in \overline{V}$ are {\em chronologically related},
$(P,F)\overline{\ll} (P',F')$, if $F\cap P'\neq\emptyset$.
Regarding to the extension of the causal relation $\leqslant$, it is
enough to consider the following sufficient criterium for $(P,F),
(P',F')\in \overline{V}$ with, either $P\neq\emptyset$ or $F'\neq
\emptyset$:
\[
P\subset P'\;\;\hbox{and}\;\; F'\subset F \Rightarrow
(P,F)\overline{\leqslant} (P',F').
\]
In the particular case of the spacetimes treated in this paper
(see Section \ref{4.2}),
this criterium can be also regarded as the general definition of
$\overline{\leqslant}$ in $\overline{V}$ (recall the discussion in
\cite[Sect. 6.4]{FHSst} and references therein). Moreover, we will
say that two different pairs in $\overline{V}$ are {\em
horismotically related} if they are causally but not
chronologically related.

Finally, the topology of the spacetime is also extended to the
c-completion by means of the so-called {\em chronological topology
} ({\em chr. topology}, for short). In order to define it, first
consider the following
{\em limit operator} $L$ for $\overline{V}$: given a sequence
$\sigma=\{(P_{n},F_{n})\}\subset\overline{V}$,
$$(P,F)\in L(\sigma)\iff\left\{ \begin{array}{ccc} P\in \hat{L}(\{P_n\}) & \hbox{whenever} &  P\neq \emptyset\\ F\in \check{L}(\{F_n\}) & \hbox{whenever} & F\neq \emptyset, \end{array}\right.$$ where
\begin{equation}\label{limcrono}
\begin{array}{c}
\hat{L}(\{P_{n}\}):=\{P'\in\hat{V}: P'\subset {\rm
LI}(\{P_{n}\})\;\;\hbox{and}\;\; P'\;\;\hbox{is a maximal IP into}\;\; {\rm LS}(\{P_{n}\})\} \\
\check{L}(\{F_{n}\}):=\{F'\in\check{V}: F'\subset {\rm
LI}(\{F_{n}\})\;\;\hbox{and}\;\; F'\;\;\hbox{is a maximal IF
into}\;\; {\rm LS}(\{F_{n}\})\}
\end{array}
\end{equation}
(LI and LS are the usual point set inferior and superior limits of
sets).
Then, one can check that a topology is defined on $\overline{V}$
as follows:
\[
\hbox{$C$ is {\em closed} $\Leftrightarrow$ $L(\sigma)\subset C$
for any sequence $\sigma\subset C$.}
\]
Note that a topology on the future (resp. past) c-completion
$\hat{V}$ (resp. $\check{V}$) can be defined in a similar way,
just by using the limit operator $\hat{L}$ (resp. $\check{L}$)
instead of $L$. In this case, the resulting topology, which also
extends the topology of the spacetime, is called the {\em future}
(resp. {\em past}) {\em chronological topology}.
\begin{remark}\label{propsimplepunt} {\em We remark
the following basic properties about the chronological topology:
\begin{itemize}
\item[(1)] The chronological topology (as well as the future and
past ones) is sequential and $T_1$ (see \cite[Prop. 3.39 and 3.21]{FHSconf}), but may
be non-Hausdorff.

\item[(2)] Clearly, if $(P,F)\in L(\{(P_{n},F_{n})\})$ then
$\{(P_{n},F_{n})\}$ converges to $(P,F)$. When the  converse
happens, $L$ is called {\em of first order} (see \cite[Section
3.6]{FHSconf}).

\item[(3)] Given a pair $(P,F)\in \partial V$, any timelike curve
defining $P$ (or $F$) converges to $(P,F)$ with the chronological
topology (see \cite[Th. 3.27]{FHSconf}).
\end{itemize}
}
\end{remark}

\smallskip

These definitions for the c-boundary construction involve some
particular 
subtleties, which are essentially associated to the following two
facts: first, a TIP (or TIF) may not determine a unique pair in
the c-boundary, and, second, the topology does not always agree
with the S-relation, in the sense that, for S-related elements as
above
$$P\in \hat{L}(P_n)\not\Leftrightarrow F\in \check{L}(F_n).$$ This
makes natural to remark the following special cases:
\begin{definition}\label{simpletop}
A spacetime $V$ has a c-completion $\overline{V}$ which is {\em
simple as a point set} if each TIP (resp. each TIF) determines a
unique pair in $\partial V$.

Moreover, the c-completion
is {\em simple} if it is simple as a point set and also {\em
topologically simple}, i.e. $(P,F)\in L(P_{n},F_{n})$ holds when
either $P\in \hat{L}(\{P_{n}\})$ or $F\in \check{L}(\{F_{n}\})$.
\end{definition}

\begin{remark}
{\em The previous definition is slightly redundant. Even though
the notion of {\em simple} comprises two levels, simplicity as a
point set and as a topological set, really it is equivalent to the
second level. In fact, if the c-completion is topologically simple
and we assume by contradiction the existence of, say,
$(P,F_1),(P,F_2)\in \partial V$ with $F_1\neq F_2$, then the
constant sequence $\{(P,F_1)\}$ converges to $(P,F_1)$ and not to
$(P,F_2)$ (as the c-completion is always $T_1$) in contradiction
with topological simplicity. }
\end{remark}

\subsection{Finsler Manifolds}


A Finsler metric $F$ on a manifold $M$ gives smoothly {\em a
positively homogeneous norm}  at each $p\in M$, where  positive
homogeneity means that the usual equality $F(\lambda v)=|\lambda|
F(v)$ for $v\in TM$ of the usual norms is assumed only when
$\lambda\geq 0$. So, given such a $F$, one can define the {\em
reverse Finsler metric}: $F^{{\rm rev}}(v) := F(-v)$. Given a
Finsler manifold
$(M,F)$, a map $d:M\times M\rightarrow \R$
is defined in the
following way:
\begin{equation}\label{e10}
d(x,y):=\inf_{c\in C(x,y)}{\rm length}(c)=\inf_{c\in
C(x,y)}\int_{s_0}^{s_1}F(\dot{c}(s))ds,
\end{equation}
where $C(x,y)$ is the set of piecewise smooth curves
$c:[s_0,s_1]\rightarrow M$ with $c(s_0)=x$, $c(s_1)=y$. Such a $d$
is a {\em generalized distance}, that is, it satisfies all the
axioms of a distance except symmetry (i.e. $d$ is a {\em
quasi-distance}) and, additionally, the following condition holds:
a sequence $\{x_n\}\subset M$ satisfies $d(x,x_n)\rightarrow 0$
iff $d(x_n,x)\rightarrow 0$. One can define the forward and
backward open balls of center $x_0\in M$ and radius
$r>0$ as: $ B^+(x_0,r)=\{x\in M: d(x_0,x)<r\}$ and $
B^-(x_0,r)=\{x\in M: d(x,x_0)<r\}$, resp. Each type of balls
constitutes a topological basis of $M$.
\vspace{2mm}

A sequence $\sigma=\{x_{n}\}$ in $(M,F)$ is a (forward) Cauchy
sequence if for all $\epsilon>0$ there exists $n_0$ such that
$d(x_n,x_m)<\epsilon$ whenever $n_0\leqslant n\leqslant m$. Let ${\rm
Cau}(M)$ be the space of all the Cauchy sequences in $(M,F)$. Two
Cauchy sequences $\sigma, \sigma'\in {\rm Cau}(M)$, with
$\sigma=\{x_{n}\}$, $\sigma'=\{x'_{n}\}$, are related
$\sigma\sim\sigma'$ iff:
$$\lim_{n}\lim_{m}d(x_n,x'_m)=\lim_{n}\lim_{m}d(x'_n,x_m)=0.$$
With these notions at hand, the (forward) Cauchy completion
$M_C^+$ and (forward) Cauchy boundary $\partial_C^+ M$ of $(M,F)$
are defined by:
$$M_C^+={\rm Cau}(M)/\sim \qquad \partial_C^+ M:=M^+_C\setminus
M.$$ The {\em backward Cauchy sequence}, its correspondence space
${\rm Cau}^{{\rm rev}}(M)$, and consequently, the {\em backward
Cauchy completion} $M_C^-$ and the {\em backward Cauchy boundary}
$\partial_C^- M$, are defined analogously but using $d^{{\rm
rev}}$, defined as $d^{{\rm rev}}(x,y):=d(y,x)$ for all $x, y$.
Note that $d^{{\rm rev}}$ can be also obtained from (\ref{e10})
by replacing $F$ by $F^{{\rm rev}}$.

The generalized distance $d$ (resp. $d^{\rm rev}$) on $M$ can be
extended to a quasi-distance $d_{Q}:M_C^+\times M_C^+\rightarrow
[0,\infty]$ (resp. $d_{Q}^{\rm rev}:M_C^-\times M_C^-\rightarrow
[0,\infty]$) in a natural way, i.e.
\[
\begin{array}{c}
d_{Q}([\{x_n\}],[\{y_n\}]):=\lim_{n}(\lim_{m}d(x_n,y_m)) \\
(\hbox{resp.}\;\;d_{Q}^{{\rm
rev}}([\{x_n\}],[\{y_n\}]):=\lim_{n}(\lim_{m}d^{{\rm
rev}}(x_n,y_m))).
\end{array}
\]
In general, as $d_Q$ (resp. $d_Q^{\rm rev}$) is {\em not} a generalized
distance (see \cite[Example 3.24]{FHSst}), it generates two (in
general, different) topologies on $M_C^+$ (resp. $M_C^-$): one by
using the corresponding forward balls and the other one by using
the backward ones. The natural one, so that {\em forward} (resp.
backward) Cauchy sequences converge to the point represented by
its class in $M_C^+$ (resp. $M_C^-$), is the generated by the {\em
backward} (resp. forward) balls.

With this choice the natural inclusion $i:M\hookrightarrow M_C^+$
(resp. $i:M\hookrightarrow M_C^-$) becomes a topological
embedding, being $\partial_C^+M$ (resp. $\partial_C^-M$) a closed
subset of $M_C^+$ (resp. $M_C^-$). The topology associated to
$d_{Q}$ (resp. $d_{Q}^{{\rm rev}}$) is not Hausdorff (neither
$T_1$) in general, but only $T_0$.
%
On the other hand, the distance $d_{Q}$ (resp. $d_{Q}^{{\rm
rev}}$) can be extended (with the same formal definition) to
$d_{Q}:M_C^+\times (M_C^+\cup M_{C}^-)\rightarrow [0,\infty]$
(resp. $d_{Q}^{{\rm rev}}:(M_C^+\cup M_C^-)\times M_C^-
\rightarrow [0,\infty]$).

\smallskip

The Cauchy boundaries $\partial_C^+M, \partial_C^-M$ can be
related thanks to the following property: if $\sigma,\sigma'\in
{\rm Cau}(M)$ are related for $d$, and $\sigma\in {\rm Cau}^{{\rm
rev}}(M)$, then $\sigma'\in {\rm Cau}^{{\rm rev}}(M)$ and
$\sigma,\sigma'$ are related for $d^{{\rm rev}}$ (\cite[Prop.
3.20]{FHSst}). So, one can define the {\em symmetrized Cauchy
boundary} as the following intersection:
\begin{equation}\label{cauchysimetrizado}\partial_C^sM=\partial_C^+M\cap \partial_C^-M.\end{equation}
Moreover, a sequence $\sigma\in \partial_C^s M$ iff it is a Cauchy
sequence (in the classical sense) for the {\em symmetrized
distance}
$$d^s(x,y):=\frac{d(x,y)+d^{{\rm rev}}(x,y)}{2},$$
which is a (true) distance, even though it cannot be obtained as a
length space in general (i.e. as the infimum of lengths of
connecting curves). On $M_{C}^{s}=M\cup \partial_C^s M$ one can
also define the extension $\overline{d}^{s}$ of $d^{s}$ on $M$,
which satisfies $\overline{d}^{s}=(d_{Q}+d_{Q}^{{\rm rev}})/2$.
All the relevant properties of these constructions are summarized
in \cite[Theor. 1.1]{FHSst}.

%

\subsection{Randers to stationary correspondence}

Taking into account  the conformal invariance of the c-boundary,
we consider, without loss of generality, the normalized expression
(\ref{inteest}) below as the definition of any standard stationary
spacetime:
\begin{equation}\label{inteest} V=(\R\times M,
g=-dt^2+\pi_M^*\omega \otimes dt + dt\otimes \pi_M^*\omega + \pi_M^*h),
\end{equation} where $\omega$ is a 1-form, $h$ is a Riemannian metric,
both on $M$, and $\pi_M:\R\times M\rightarrow M$, $t:\R\times
M\rightarrow \R$ are the natural projections. The elements in
(\ref{inteest}) allow to construct the following Finsler metrics
(of Randers type) on $M$:
\[
\begin{array}{l}
F^+(v)(=F(v))=\sqrt{h(v,v)+\omega(v)^2}+ \omega(v), \\
F^-(v)(=F^{{\rm rev}}(v))=\sqrt{h(v,v)+\omega(v)^2}- \omega(v).
\end{array}
\]
We will denote by ${\rm length}_{\pm}$ and $d^\pm$, resp., the
generalized length and distance associated to $F^\pm$ (as in
(\ref{e10})), and by $d_{Q}^{\pm}$ the natural extension of
$d^\pm$ to $M_{C}^{\pm}$.

In \cite{CJS} the authors analyzed the relation between different
properties of stationary spacetimes and these Finsler metrics of
Randers type. In \cite{FHSst} this study was extended to the
c-boundary. In the next section we revisit this second study by
reproducing it in a slightly more general framework.

%

\section{Revisiting the c-boundary of stationary
spacetimes} \label{s3}

\label{ss1.3}

Consider a connected Riemannian manifold $(M,h)$, a $1$-form
$\omega$ and the natural projections $\pi_M,t$ as above. For each
smooth positive function $\tilde{\alpha}:\R\rightarrow\R$,
consider the spacetime $V_{\tilde\alpha}:=(\R\times
M,g_{\tilde\alpha})$, with
\begin{equation}\label{eee}
g_{\tilde{\alpha}}:=-dt^2+\replace{\sqrt{\tilde{\alpha}(t)}}{\tilde{\alpha}(t)}\pi_M^*\omega\otimes
dt+\replace{\sqrt{\tilde{\alpha}(t)}}{\tilde{\alpha}(t)} dt\otimes\pi_M^*\omega
+\replace{\tilde{\alpha}}{\tilde{\alpha}^2}(t) \pi_M^*h,
\end{equation}
where the pull-back $\pi_M^*$ will be omitted when there is no
possibility of confusion and $\tilde\alpha (t)$ denotes $
\tilde\alpha \circ t$ (the future time-orientation is always
determined by $\partial_t$). Up to the conformal factor
$\replace{\tilde{\alpha}}{\tilde{\alpha}^2}(t)$, the change of
variables
$ds=dt/\replace{\sqrt{\tilde{\alpha}(t)}}{\tilde{\alpha}(t)}$
allows to identify the spacetime $V_{\tilde{\alpha}}$ with a
region $I\times M$ ($I\subset \R$ interval, $0\in I$) of the
standard stationary spacetime (\ref{inteest}).
Moreover, this region $I\times M$ becomes the full standard
stationary spacetime (i.e., $I=\R$) if and only if the following
condition holds (see the proof of Proposition \ref{auxprop} for
details):
\begin{equation}\label{intcond}
\int_{t=0}^{t=\infty}\frac{ds}{\replace{\sqrt{\tilde{\alpha}(s)}}{\talpha(s)}}
=\int_ {t=-\infty}^{t=0}\frac{ds}{\replace{\sqrt{\tilde{\alpha}}(s)}{\talpha(s)}}=\infty.
\end{equation}
The results in \cite[Sect. 6]{FHSst} determine accurately the
c-boundary of  any standard stationary spacetime, and thus, under
condition (\ref{intcond}), the c-boundary of the spacetime
$V_{\tilde\alpha}$ as well. Next, we will translate the results in
this reference to describe the c-boundary of $V_{\tilde\alpha}$
maintaining explicitly the dependence on $\tilde{\alpha}$ (which
will be useful for the isocausal comparison). We will assume
(\ref{intcond}) throughout all the paper, even though the steps in
Subsection \ref{ff} do not require it.



\subsection{Relation between chronology and Busemann
functions}\label{ff}

Let us begin by establishing the following characterization of the
chronological relation in $V_{\tilde{\alpha}}$ (see \cite[Prop.
6.1]{FHSst}):
\begin{proposition}\label{p0}
If $\gamma(t)=(t,c(t))$, $t\in [a,\Omega)$, is a future-directed
timelike curve in $V_{\tilde{\alpha}}=(\R\times
M,g_{\tilde{\alpha}})$. Then $F^+(\dot{c}(t))<1/\talpha (t)$ and ${\rm
length}_{+}(c\mid_{[t_{0},t_{1}]})<\int_{t_0}^{t_1}ds/\replace{\sqrt{\tilde{
\alpha}}(s)}{\talpha(s)}$, with $\alpha\leqslant t_0\leqslant t_1<\Omega$.
Moreover,
\[
(t_{0},x_{0})\ll_{\tilde{\alpha}}(t_{1},x_{1}) \Longleftrightarrow
d^{+}(x_{0},x_{1})<\int_{t_0}^{t_1}\frac{ds}{\replace{\sqrt{\tilde{\alpha}}(s)}{
\talpha(s)}}.
\]
\end{proposition}
As a direct consequence, note that $\int_{t_0}^{t}ds/\talpha(s) -
d(\cdot,c(t))$ increases with $t$.

\smallskip

According to Section \ref{d}, any IP in $(\R\times
M,g_{\tilde{\alpha}})$ is of the form
$P_{\tilde{\alpha}}=I^-_{\tilde{\alpha}}(\gamma)$, being $\gamma$
a future-directed timelike curve with $\gamma(t)=(t,c(t))$ and
$t\in [a,\Omega)$. Then, from Proposition \ref{p0} one
deduces
\begin{equation}\label{defbuse}
\begin{array}{rl}P_{\tilde{\alpha}}= & I^{-}_{\tilde{\alpha}}(\gamma) \\ = & \{(t_{0},x_{0})\in V_{\tilde{\alpha}}: (t_{0},x_{0})\ll_{\tilde{\alpha}}(t,c(t))\;\;\hbox{for some
$t$ close enough  to $\Omega$}\} \\  = & \{(t_0,x_0)\in
V_{\tilde{\alpha}}:
0<\lim_{t\rightarrow\Omega}(\int_{t_{0}}^{t}\frac{ds}{\replace{\sqrt{\tilde{
\alpha}(s)}}{\talpha(s)}}-d^{+}(x_{0},c(t)))\}
\\ = & \{(t_0,x_0)\in V_{\tilde{\alpha}}: \int_{0}^{t_0}
\frac{ds}{\replace{\sqrt{\tilde{\alpha}(s)}}{\talpha(s)}}<\lim_{t\rightarrow
\Omega}(\int_{0}^t
\frac{ds}{\replace{\sqrt{\tilde{\alpha}}}{\talpha}(s)}-d^+(x_0,c(t)))\}.
\end{array}
\end{equation}
Therefore, if one defines the {\em forward Busemann function} for
$g_{\tilde{\alpha}}$ associated to a curve $c$ with $F^{+}(\dot{c}(t))<1/\tilde{\alpha}(t)$ as
\begin{equation}\label{busemanfunc}
b_{c,{\tilde{\alpha}}}^+(\cdot):=\lim_{t\rightarrow
\Omega}\left(\int_{0}^t
\frac{ds}{\replace{\sqrt{\tilde{\alpha}(s)}}{\talpha(s)}}-d^+(\cdot,
c(t))\right) ,
\end{equation}
one can write
$P_{\tilde{\alpha}}=P_{\tilde{\alpha}}(b_{c,{\tilde{\alpha}}}^+)$
where, for any function $f$,
\begin{equation}\label{defbuse'}
P_{\tilde{\alpha}}(f) =\left\{(t_0,x_0)\in V_{\tilde{\alpha}}:
\int_{0}^{t_0}
\frac{ds}{\replace{\sqrt{\tilde{\alpha}(s)}}{\talpha(s)}}<f(x_0)\right\}.
\end{equation} Similarly, any IF in $V_{\tilde{\alpha}}$ is
of the form $F_{\tilde{\alpha}}=I^+_{\tilde{\alpha}}(\gamma)$,
being $\gamma$ a past-directed timelike curve with
$\gamma(t)=(-t,c(t))$ and $t\in [a,-\Omega)$. Thus,
$F_{\tilde{\alpha}}=F_{\tilde{\alpha}}(b_{c,{\tilde{\alpha}}}^-)$
where
\begin{equation}\label{defbuse''}
F_{\tilde{\alpha}}(f) =\left\{(t_0,x_0)\in V_{\tilde{\alpha}}:
\int_{0}^{t_0}
\frac{ds}{\replace{\sqrt{\tilde{\alpha}(s)}}{\tilde{\alpha}(s)}}
>f(x_0)\right\},
\end{equation}
and
\[
b_{c,{\tilde{\alpha}}}^-(\cdot):=\lim_{t\rightarrow
-\Omega}\left(-\int_{-t}^{0}
\frac{ds}{\replace{\sqrt{\tilde{\alpha}(s)}}{\tilde{\alpha}(s)}}+d^-(\cdot,
c(t))\right)=
\lim_{t\rightarrow
-\Omega}\left(\int_{0}^{-t}
\frac{ds}{\replace{\sqrt{\tilde{\alpha}(s)}}{\tilde{\alpha}(s)}}+d^-(\cdot,
c(t))\right)
\]
is the {\em backward Busemann function} for $g_{\tilde{\alpha}}$
associated to a curve $c$ with
$F^{-}(\dot{c}(t))<1/\tilde{\alpha}(t)$. In conclusion, the set of
all IPs (resp. IFs) can be identified with the set of all forward
(resp. backward) Busemann functions. So, denoting as
$B_{\tilde{\alpha}}^+(M)$ (resp. $B_{\tilde{\alpha}}^-(M)$) the
set of all the finite forward (resp. backward) Busemann functions,
one has:\footnote{Note that a forward (resp. backward) Busemann
function takes the value plus (resp. minus) infinity at some point
iff it is constantly equal to plus (resp. minus) infinity.}
\[
\hat{V}_{\tilde{\alpha}}\equiv
B^{+}_{\tilde{\alpha}}(M)\cup\{\infty\}\qquad
\check{V}_{\tilde{\alpha}}\equiv
B^{-}_{\tilde{\alpha}}(M)\cup\{-\infty\}.
\]
Next, assume $\pm\Omega<\infty$. From Prop. \ref{p0}, ${\rm
length}_{\pm}(c|_{[a,\pm\Omega)})<\int_{a}^{\pm
\Omega}ds/\replace{\sqrt{\tilde{\alpha}(s)}}{\tilde{\alpha}(s)}<\infty$, and
so, $c(t)\rightarrow
x^\pm$ for some $x^\pm\in M_{C}^\pm$. In particular, for each $\Omega\in \R$, putting $\Omega_{\tilde
\alpha}:=\int_0^{\Omega}ds/\replace{\sqrt{\tilde{\alpha}(s)}}{\tilde{\alpha}(s)}
$:
\begin{equation}\label{ddd2}
b_{c,{\tilde{\alpha}}}^\pm(\cdot)=\int_{0}^{\Omega}\frac{ds}{\replace{\sqrt{
\tilde{\alpha}(s)}}{\tilde{\alpha}(s)}}\mp
d_Q^\pm(\cdot,x^\pm)=\Omega_{\tilde{\alpha}}\mp
d_Q^\pm(\cdot,x^\pm),\quad x^{\pm}\in M_{C}^{\pm}.
\end{equation}
Therefore, if we define
\begin{equation}\label{ddd1}d^\pm_{(\Omega_{\tilde{\alpha}},x^\pm)}(\cdot):=\Omega_{\tilde{\alpha}}\mp
d_Q^\pm(\cdot,x^\pm)\quad \hbox{for any $x^{\pm}\in
M_{C}^{\pm}$},\end{equation} we have just deduced that
\begin{equation}\label{ddd}
b_{c,{\tilde{\alpha}}}^\pm(\cdot)=d^\pm_{(\Omega_{\tilde{\alpha}},x^\pm)}(\cdot)\qquad\hbox{if}\quad
\pm\Omega<\infty.
\end{equation}

\subsection{Point set structure of the partial boundaries}

In the remainder of this section, we will describe the c-boundary
of $V_{\tilde\alpha}$ for $I=\R$, i.e., under condition
(\ref{intcond}). So, the following action on
$B^\pm_{\tilde{\alpha}}(M)$ is well-defined:
\begin{equation}\label{ac}\begin{array}{ccc} B^\pm_{\tilde{\alpha}}(M) \times \R & \rightarrow & B^\pm_{\tilde{\alpha}}(M) \\ (b_{c,\tilde{\alpha}}^\pm,K) &
\mapsto & b_{c,\tilde{\alpha}}^\pm + K.
\end{array}\end{equation}
In fact, (for the case $^+$) standard arguments allow one to
deduce from condition (\ref{intcond}) the existence of some
increasing bijective smooth function $s_K(t)$ such that:
\[
\int_{t}^{s_K(t)}\frac{1}{\talpha(r)}dr=K,\qquad\hbox{and
so}\quad\;\dot{s}_{K}(t)=\frac{\talpha(s_K(t))}{\talpha(t)}.
\]
As a consequence,
\begin{equation}\label{e11}
\begin{array}{rl}
b_{c,\tilde{\alpha}}^+ + K &
=\lim_{t\rightarrow\infty}(\int_{0}^{t}\frac{dr}{\replace{\sqrt{\tilde{\alpha}
(r)}}{\tilde{\alpha}(r)}}-d^{+}(\cdot,c(t)))+K \\
&
=\lim_{t\rightarrow\infty}(\int_{0}^{t}\frac{dr}{\replace{\sqrt{\tilde{\alpha}
(r)}}{\tilde{\alpha}(r)}}+\int_{t}^{s_K(t)}\frac{dr}{\replace{\sqrt{\tilde{
\alpha}(r)}}{\tilde{\alpha}(r)}} -d^{+}(\cdot,c(t)))
\\
&
=\lim_{t\rightarrow\infty}(\int_{0}^{s_K(t)}\frac{dr}{\replace{\sqrt{\tilde{
\alpha}(r)}}{\tilde{\alpha}(r)}}-d^{+}(\cdot,c(t)))
\\ &
=\lim_{s\rightarrow\infty}(\int_{0}^{s}\frac{dr}{\replace{\sqrt{\tilde{\alpha}
(r)}}{\tilde{\alpha}(r)}}-d^{+}(\cdot,\overline{c}(s))),\quad\hbox{where
$ \overline{c}(s_K(t)):=c(t)$,}
\end{array}
\end{equation}
and notice that
\begin{equation}\label{nuevo}
F(\dot{\overline{c}}(s_K(t)))=F(\dot{c}(t)/\dot{s}_K(t))=\frac{F(\dot{c}(t))}{\dot{s}_K(t)}<\frac{1}{\talpha(t)}\cdot\frac{\talpha(t)}{\talpha(s_K(t))}=\frac{1}{\talpha(s_K(t))}.
\end{equation}
So, the last term in (\ref{e11}) (under inequality (\ref{nuevo})
for $\overline{c}$) is just the definition of
$b^{+}_{\overline{c},\tilde{\alpha}}\in
B^{+}_{\tilde{\alpha}}(M)$, as required.

%

Next, define the {\em forward/backward Busemann completion} as the
quotient space
$M_{B,\tilde{\alpha}}^\pm=B_{\tilde{\alpha}}^\pm(M)/\R$ provided
by the action (\ref{ac}). From (\ref{ddd}), the forward/backward
Busemann completion $M_{B,\tilde{\alpha}}^\pm$ clearly contains
the forward/backward Cauchy completion $M_{C}^{\pm}$. So, if we
define the {\em forward/backward Busemann boundary} as
$\partial_{B,{\tilde{\alpha}}}^{\pm}M:=M_{B,{\tilde{\alpha}}}^\pm\setminus
M$ and the {\em forward/backward proper Busemann boundary} as
$\partial_{{\cal
B},{\tilde{\alpha}}}^{\pm}M:=\partial_{B,{\tilde{\alpha}}}^{\pm}M\setminus
\partial_{C}^{\pm}M$, one obtains the decomposition:
\[
M_{B,{\tilde{\alpha}}}^\pm=M\cup
\partial_{B,{\tilde{\alpha}}}^{\pm}M=M\cup
\partial_C^\pm M\cup \partial_{{\cal B},{\tilde{\alpha}}}^\pm M.
\]
Finally, labelling the TIP (resp. TIF) $V=\R\times M$ by $i^{+}$
(resp. $i^{-}$), one can write (see \cite[Sect. 6.2]{FHSst}):
\begin{equation}\label{strucbor}\begin{array}{rl}\hat{V}_{\tilde{\alpha}}\equiv &  B^+_{\tilde{\alpha}}(M)\cup \{\infty\} = (\R\times M_{B,{\tilde{\alpha}}}^+)\cup \{i^{+}\} \\ =& (\R\times (M\cup\partial_{B,\tilde{\alpha}}^{+}M))\cup \{i^+\}\\ = & V\cup (\R\times\partial_{B,\tilde{\alpha}}^{+}M)\cup \{i^{+}\} \\ = & V\cup (\R\times\partial_C^+ M)\cup (\R\times\partial_{{\cal B},{\tilde{\alpha}}}^+M)\cup \{i^{+}\}.\\ & \\\hat{\partial}_{\tilde{\alpha}}V
\equiv &(\R\times\partial^{+}_{B,{\tilde{\alpha}}} M)\cup
\{i^{+}\}= (\R\times\partial_{C}^{+}M)\cup
(\R\times\partial_{{\cal B},{\tilde{\alpha}}}^{+}M)\cup
\{i^{+}\}.\\ & \\ \check{V}_{\tilde{\alpha}}\equiv & V\cup
(\R\times\partial_C^-
M)\cup (\R\times\partial_{{\cal B},{\tilde{\alpha}}}^-M)\cup \{i^{-}\}.\\
&
\\\check{\partial}_{\tilde{\alpha}}V \equiv& (\R\times\partial_{C}^{-}M)\cup
(\R\times\partial_{{\cal B},{\tilde{\alpha}}}^{-}M)\cup
\{i^{-}\}.\end{array}
\end{equation}
%
%

\subsection{Topology and S-relations}\label{s3.3}

The future and past chronological topologies can be characterized
by using the identifications (\ref{strucbor}) in terms of the
following limit operators, whose dependence on $\tilde{\alpha}$,
via their domain, is implicitly assumed: given a sequence
$\{f_{n}\}_{n}$ and an element $f$ in
$B^{\pm}_{\tilde{\alpha}}(M)\cup\{\pm\infty\}$,\footnote{The term
$\{\pm\infty\}$ is included here in order to obtain a more direct
relation with the c-boundary of spacetimes. Note that this term is
not necessary when one works with Busemann compactifications,  as
a quotient space by additive constants is considered then (compare
with \cite{FHSst}).}
\[
\begin{array}{l}f\in \hat{L}(\{f_n\})\iff \left\{\begin{array}{l}(a)\; f\leqslant
\liminf_{n}f_{n} \hbox{ and}\\ (b)\; \forall g\in
B^{+}_{\tilde{\alpha}}(M) \hbox{ with } f\leqslant g\leqslant \limsup_{n}
f_{n}, \hbox{ it is } g=f\end{array}\right.
\\ \\f\in \check{L}(\{f_n\})\iff \left\{\begin{array}{l} (a)\; f\geq
\limsup_{n}f_{n} \hbox{ and}\\ (b)\; \forall g\in
B^{-}_{\tilde{\alpha}}(M) \hbox{ with } f\geq g\geq \liminf_{n}
f_{n}, \hbox{ it is } g=f.
\end{array}\right.\end{array}
\]

The future and past chronological topologies are coarser than the
pointwise topology in the space of Busemann functions. In fact, we
have the following adaptation of \cite[Proposition 5.29]{FHSst}.
\begin{proposition}\label{proposition1}
 Consider a sequence $\{f_n\}\subset B^\pm_{\bar{\alpha}}(M)$ which
converges pointwise to a function $f\in B^\pm_{\bar{\alpha}}(M)$.
Then, $f$ is the unique future (resp. past) chronological limit of
$\{f_n\}$.

In particular, if $f_{n}=d^\pm_{(\Omega^n_{\bar{\alpha}},x^n)}$
with $\Omega^n_{\bar{\alpha}}\rightarrow \Omega_{\bar{\alpha}}(\in \R)$
and $x^n\rightarrow x(\in M_C^\pm)$, then $f=d^+_{(\Omega_{\bar{\alpha}},x)}\in
\hat{L}(f_n)$ (resp. $f=d^-_{(\Omega_{\bar{\alpha}},x)}\in
\check{L}(f_n)$) is the unique future (resp. past) chronological
limit of $\{f_n\}$.
\end{proposition}

In order to study the S-relations (\ref{eSz}), (\ref{eSz2})
between terminal sets, their common pasts and futures must be
considered. If $P_{\tilde{\alpha}}$ is generated by a
future-directed timelike curve $\gamma (t)=(t,c(t))$, $t\in
[a,\infty)$, then $\uparrow_{\tilde{\alpha}}
P_{\tilde{\alpha}}=\emptyset$, and so, $P_{\tilde{\alpha}}$ is
S-related only to $\emptyset$. Analogously, if
$F_{\tilde{\alpha}}$ is generated by a past-directed timelike
curve $\gamma(t)=(-t,c(t))$, $t\in [-a,\infty)$, then
$\downarrow_{\tilde{\alpha}} F_{\tilde{\alpha}}=\emptyset$, and
$F_{\tilde{\alpha}}$ is S-related only to $\emptyset$. If,
instead, $\pm\Omega<\infty$, then
$b_{c,\tilde{\alpha}}^{\pm}=d^{\pm}_{p^\pm}$, with
$p^\pm=(\Omega_{\tilde{\alpha}},x^\pm)$ (recall (\ref{ddd})) and,
thus:
\begin{equation}\label{srelation}\begin{array}{l}\uparrow P_{\tilde{\alpha}}=\uparrow
P_{\tilde{\alpha}}(d^+_{p^+})=F_{\tilde{\alpha}}(d^-_{p^+})\\
\downarrow F_{\tilde{\alpha}}=\downarrow
F_{\tilde{\alpha}}(d_{p^-}^-)=P_{\tilde{\alpha}}(d_{p^-}^+).\end{array}\end{equation}
For the first equality of each line in (\ref{srelation}) recall
Section \ref{ff},
and for the second one recall \cite[Lemma 6.11]{FHSst}.

\begin{remark}\label{remark1}
{\em Previous expressions allow to deduce the following facts: (i) if
$p=p^+=p^-=(\Omega_{\tilde{\alpha}},x^s)\in\R\times\partial_C^s M$
(i.e., the future-directed and past-directed timelike curves
generating the terminal sets end at the same point of $\R\times
\partial_C^s M$, recall (\ref{cauchysimetrizado})), then $\uparrow
P_{\tilde{\alpha}}(d^+_{p})=F_{\tilde{\alpha}}(d^-_{p})$,
$\downarrow
F_{\tilde{\alpha}}(d_{p}^-)=P_{\tilde{\alpha}}(d_{p}^+)$, and so,
$P_{\tilde{\alpha}}(d^+_p)\sim_S F_{\tilde{\alpha}}(d^-_p)$; (ii)
if $d_{Q}^{+}$ is not a generalized distance, but $(M,d^+)$ has an
{\em evenly pairing boundary} (see \cite[Def. 3.34]{FHSst}), then
$d^\mp_{p^\pm}=\Omega_{\tilde{\alpha}}\pm
d^{\mp}(\cdot,x^{\pm})=\pm \infty$ whenever
$x^\pm\in\partial_C^\pm M\setminus\partial_C^s M$ and, in this
case, $\uparrow P_{\tilde{\alpha}}(d^+_{p^+})=\emptyset=\downarrow
F_{\tilde{\alpha}}(d^-_{p^-})$ (for non evenly pairing boundaries,
the situation is much subtler, see \cite[Th 1.2
(1B)(c2)]{FHSst}).}
\end{remark}

\subsection{Main result}

The action (\ref{ac}) defined on Busemann functions (always under
condition (\ref{intcond})) induces also a natural action on
$S$-related pairs, each one generating the following orbit ({\em
line}):
\[\begin{array}{l}\Lin(P_{\tilde{\alpha}},\emptyset):=\{(P'_{\tilde{\alpha}},
\emptyset):P'_{\tilde{\alpha}}=P_{\tilde{\alpha}}(b_{c,\alpha}^+ +k),\; k\in
\R\}\;\; \hbox{where $P_{\tilde{\alpha}}=P_{\tilde{\alpha}}(b_{c,\tilde{\alpha}}^+)$}\\
\Lin(\emptyset,F_{\tilde{\alpha}}):=\{(\emptyset,F'_{\tilde{\alpha}}):F'_{
\tilde{\alpha}}=F_{\tilde{\alpha}}(b_{c,\tilde{\alpha}}^- +k),\;
k\in \R\}\;\; \hbox{where $F_{\tilde{\alpha}}=F_{\tilde{\alpha}}(b_{c,\tilde{\alpha}}^-)$}
\end{array}
\]
and
\[
\Lin(P_{\tilde{\alpha}},F_{\tilde{\alpha}}):=\{(P_{\tilde{\alpha}}',F'_{
\tilde{\alpha}}):P'_{\tilde{\alpha}}=P_{\tilde{\alpha}}(d_p^+ +k),
F'_{\tilde{\alpha}}=F_{\tilde{\alpha}}(d_{p'}^- +k), k\in \R\},
\]
where $P_{\tilde{\alpha}}=P_{\tilde{\alpha}}(d_p^+)$ and
$F_{\tilde{\alpha}}=F_{\tilde{\alpha}}(d_{p'}^-)$. Now, Theorem
1.2 in \cite{FHSst} yields, in particular, the following
result.
\begin{theorem}\label{theo1}
Let $V_{\tilde{\alpha}}=(\R\times M,g_{\tilde{\alpha}})$ be a
spacetime as in (\ref{eee}) satisfying the overall condition
(\ref{intcond}). If $d_Q^+$ is a generalized distance,
then the c-boundary $\partial_{\tilde{\alpha}} V$ has the
following structures.

Point Set:\begin{itemize}\item The future (resp. past) c-boundary
$\hat{\partial}_{\tilde{\alpha}} V$ (resp.
$\check{\partial}_{\tilde{\alpha}} V$) is naturally a point set
cone with base $\partial_{B,\tilde{\alpha}}^+ M$ (resp.
$\partial_{B,\tilde{\alpha}}^- M$) and apex $i^+$ (resp.
$i^-$).\item The total c-boundary $\partial_{\tilde{\alpha}}V$ is
the quotient set of the partial boundaries
$\hat{\partial}_{\tilde{\alpha}} V$,
$\check{\partial}_{\tilde{\alpha}} V$ under the S-relation. So, it
is composed by the set of all the lines
$\Lin((P_{\tilde{\alpha}}(d_p^+),F_{\tilde{\alpha}}(d_p^-)))$,
$\Lin((P_{\tilde{\alpha}}(b_{c,\tilde{\alpha}}^+),\emptyset))$,
$\Lin((\emptyset,F_{\tilde{\alpha}}(b_{c,\tilde{\alpha}}^-)))$, with
$p\in \R\times\partial_C^s M$, $b_{c,\tilde{\alpha}}^\pm\in {\cal
B}^\pm_{\tilde{\alpha}} (M)$.\end{itemize}

Causality: \begin{itemize} \item The lines
$\Lin((P_{\tilde{\alpha}}(d_p^+),F_{\tilde{\alpha}}(d_p^-)))$,
with $p\in \R\times\partial_C^s M$, are timelike, i.e. any two
points of the line are chronologically related. \item The lines
$\Lin((P_{\tilde{\alpha}}(b_{c_,\tilde{\alpha}}^+),\emptyset))$,
$\Lin((\emptyset,F_{\tilde{\alpha}}(b_{c,\tilde{\alpha}}^-))$,
with $b_{c,\tilde{\alpha}}^\pm\in {\cal
B}^\pm_{\tilde{\alpha}}(M)$, are horismotic, i.e. any two points
of the lines are causally but not chronologically related.
\end{itemize}

Topology:
\begin{itemize}\item If
$M_{B,\tilde{\alpha}}^+$ (resp. $M_{B,\tilde{\alpha}}^-$) is
Hausdorff, the future (resp. past) c-boundary has the structure of
a (topological) cone with base $\partial_{B,\tilde{\alpha}}^+M$
(resp. $\partial_{B,\tilde{\alpha}}^-M$) and apex $i^+$ (resp.
$i^-$). \item If $M_{C,\tilde{\alpha}}^{s}$ is locally compact then
$\overline{V}_{\tilde{\alpha}}$ is (topologically) simple, that
is, each TIP (resp. each TIF) determines a unique pair in
$\partial_{\talpha} V$ and $(P,F)\in L(\{(P_{n},F_{n})\})$ when
either $P\in \hat{L}(\{P_{n}\})$ or $F\in \check{L}(\{F_{n}\})$.
Equivalently, $\overline{V}_{\talpha}$ coincides with the quotient
topological space of the future and past c-completions
$\hat{V}_{\tilde{\alpha}}$ and $\check{V}_{\tilde{\alpha}}$ under
the S-relation.
\end{itemize}
\end{theorem}


\section{Framework for isocausal comparison}
\label{s4}


\subsection{Causal mappings and c-boundary}

Let us begin by recalling the notion of isocausality (see
\cite{GpScqg03} for further details):
\begin{definition}\label{defiso}
Let $V\equiv (V,g)$, $V'\equiv (V',g')$ be two
spacetimes. A diffeomorphism $\Phi: V\rightarrow V'$ is called a
{\em causal mapping} if all the future causal vectors of $V$ are
mapped by the differerential $d\Phi$ into future causal vectors of
$V'$. In this case, we write $V\prec_{\Phi}V'$. $V$ is {\em causally related} to
$V'$, denoted $V\prec
V'$, if $V\prec_{\Phi}V'$ for some causal mapping $\Phi$;
moreover, $V$ is {\em isocausal} to $V'$ if $V\prec V'$ and
$V'\prec V$.

In particular, when the underlying manifolds coincide and the identity
$Id:V\rightarrow V'$ is a causal mapping (namely,  $g(v,v)\leqslant 0$
implies $g'(v,v)\leqslant 0$) we say that the cones of $g'$ are {\em wider} than
those of $g$ and write $V\prec_0V'$ or, simply,
$g\prec_0 g'$.
\end{definition}

\begin{remark}\label{rrr}{\em Obviously, $V\prec_\Phi V'$ plus
$V'\prec_\Phi^{-1} V$ imply not only that $V$ and $V'$ are
isocausal but also that $\Phi$ is a conformal map; so, $g$ and
$g'$ are {\em conformal} metrics. If $g\prec_0 g'$ and $g'\prec_0
g$ then $g$ and $g'$ are {\em pointwise conformal} metrics.

Typically, we will take three Lorentzian metrics $g, g', g''$ on
the same manifold which satisfy $g\prec_{0} g'\prec_{0} g''$ and
use the following obvious result: if $g$ and $g''$ are conformal then $g$
and $g'$ (and $g''$) are isocausal. }
\end{remark}

\begin{remark}\label{rdis}{\em
Let us discuss the relation between the c-boundary points of two
causally related spacetimes; by taking pull-back metrics we can
consider the case $g\prec_{0} g'$ with no loss of generality.

If $P$ is a TIP for $g$, then $I^{-}_{g'}(P)$ is also a TIP for
$g'$ (write $P=I^-_g(\gamma)$ and notice
$I^{-}_{g'}(P)=I^{-}_{g'}(\gamma)$). This may suggest a procedure
to include the c-boundary for $g$ into the c-boundary for $g'$,
i.e. consider the map $P\mapsto I^-_{g'}(P)$. However, this
procedure is not satisfactory in general, as there may exist two
different TIPs $P_{1}\neq P_{2}$ for $g$ such that
$I^{-}_{g'}(P_{1})=I^{-}_{g'}(P_{2})$ (violating the injectivity
of the inclusion). This undesirable property may happen {\em even
if $g$ and $g'$ are isocausal} (see Example \ref{ex3}).

However, as we will see in the following sections, there are certain conditions under which
this procedure succeed.}\end{remark}

\subsection{Spacetimes isocausal to standard stationary ones}\label{4.2}

Our aim is to study the c-boundary of the spacetime
\begin{equation}\label{e1}
V=(\R\times M,g), \qquad g=-dt^{2}+\omega_{t}\otimes
dt+dt\otimes\omega_{t}+h_{t},
\end{equation}
where $\omega_t$ is a one form on $\R\times M$ such that
$\omega_t(\partial/\partial t)=0$ (when necessary, and abusing of
notation, it will be regarded as a one form on each slice
$\{t\}\times M$). This will be carried out by comparing it with
the c-boundaries of the spacetimes
\begin{equation}\label{exp1}
\begin{array}{l}
V_{\cc}=(\R\times M,g_{\cc}),\quad
g_{\cc}(\equiv g_{\tilde{\alpha}=1})=-dt^2+\omega\otimes dt+
dt\otimes\omega + h \\
V_{\ca}=(\R\times M,g_{\ca}),\quad
g_{\ca}(\equiv
g_{\tilde{\alpha}=\alpha})=-dt^2+\replace{\sqrt{\alpha(t)}}{\alpha(t)}
\omega\otimes dt+\replace{\sqrt{\alpha(t)}}{\alpha(t)} dt\otimes\omega
+\replace{\alpha(t)}{\alpha^2(t)} h,
\end{array}
\end{equation}
where one implicitly assumes the relations
\begin{equation}\label{exp2}
g_{cl}\prec_{0} g\prec_{0} g_{op}.
\end{equation}

First, it is convenient to rewrite condition (\ref{exp2}) in terms of the
corresponding metric coefficients. To this aim, let
$\gamma(s)=(t(s),c(s))$ be a future/past-directed timelike curve
for $g$. Then:
\[-\dot{t}^2+2\omega_t(\dot{c})\dot{t}+h_t(\dot{c},\dot{c})<0.
\]
Solving this second-order equation in $\dot{t}$, and taking into
account that $\pm\dot{t}>0$, one directly deduces:
\begin{equation}\label{charcau}
\pm\dot{t}>F_{t}^\pm(\dot{c})\qquad\hbox{with}\quad
F_{t}^\pm(\dot{c}):=\sqrt{\omega_t^2(\dot{c})+h_t(\dot{c},\dot{c})}\pm\omega_t(\dot{c}).
\end{equation}
So, the timelike character of $\gamma$ for the metric $g$ is
determined by the functions $F_{t}^\pm$. Reasoning analogously, if
$\gamma(s)=(t(s),c(s))$ is a future/past-directed timelike curve for
$g_{\cc}$ (resp. $g_{\ca}$) then
\begin{equation}\label{charcau'}
\pm\dot{t}>F^\pm(\dot{c})\qquad(\hbox{resp.}\;\;\pm
\dot{t}>\replace{\sqrt{\alpha(t)}}{\alpha(t)}F^\pm(\dot{c }
))\qquad\hbox{with}\;\;F^\pm(\dot{c}):=\sqrt{\omega^2(\dot{c})+h(\dot{c},\dot{c}
)}\pm\omega(\dot{c}).
\end{equation}

\begin{remark}\label{remcharcau}{\em
Summarizing, the causal character of the future/past directed
curve $\gamma(s)=(t(s),c(s))$ for $g,g_{\cc}$ or $g_{\ca}$, is
characterized as follows:
\[
 \begin{array}{llrl}
 \hbox{$\gamma$ is timelike (causal; lightlike) for $g$} & \hbox{if} &
F^\pm_{t}(\dot{c}) &< (\leqslant; =)\;\pm\dot{t}\\
\hbox{$\gamma$ is timelike (causal; lightlike) for $g_{\cc}$} & \hbox{if} &
F^\pm(\dot{c}) &< (\leqslant; =)\;\pm\dot{t}\\
\hbox{$\gamma$ is timelike (causal; lightlike) for $g_{\ca}$} & \hbox{if} &
\replace{\sqrt{\alpha(t)}}{\alpha(t)}F^\pm(\dot{c}) & <  (\leqslant; =)\;\pm\dot{t}\\
\end{array}
\]
}
\end{remark}
In conclusion, applying (\ref{charcau}) to tangent vectors, one deduces directly:
\begin{proposition}\label{isostat'} The relations $g_{\cc}\prec_{0} g\prec_{0} g_{op}$ hold iff the following
inequalities hold:
\begin{equation}\label{isostat}
\replace{\sqrt{\alpha(t)}}{\alpha(t)} F^\pm(\cdot)\leqslant F^\pm_t(\cdot)\leqslant
F^\pm(\cdot).
\end{equation}
\end{proposition}
On the other hand:
\begin{proposition}\label{auxprop}
$g_{\ca}$ is conformal to $g_{\cc}$ iff (\ref{intcond})
holds for $\tilde{\alpha}=\alpha$.
\end{proposition}
{\it Proof.} For the implication to the right, assume, for
example,
\[
\int_{0}^{\infty}\frac{ds}{\replace{\sqrt{\alpha(s)}}{\alpha(s)}}<\infty.
\]
From Proposition \ref{p0} with $\tilde{\alpha}\equiv \alpha$, no
timelike curve $\gamma$ for $g_{\ca}$ satisfies
$I^-_{\ca}(\gamma)=V$, as there will always exist some point with
temporal component close enough to $\Omega$ (the upper limit of
the temporal component of $\gamma$) which does not belong to the
chronological past of $\gamma$. Hence, $i^{+}\equiv V$ does not
belong to the future c-boundary of $V_{\ca}$. Nevertheless, it is
easy to check that $i^{+}$ belongs to the future c-boundary of
$V_{\cc}$ (see \cite{FHSst} or recall that $i^+$ is the past of
any curve type $t\mapsto (t,x_0)$). Therefore, the future
c-boundaries of $V_{op}$ and $V_{cl}$ cannot be identified, and
so, $g_{op}$, $g_{cl}$ are not conformally related.

For the implication to the left, note that, under (\ref{intcond})
with $\tilde{\alpha}=\alpha$, the change of variables
$ds=dt/\replace{\sqrt{\alpha(t)}}{\alpha(t)}$ takes $V_{\ca}$ into
a spacetime conformal to $V_{\cc}$. \cvd

\begin{corollary}\label{cisocausal} If the relations  $g_{cl}\prec_{0} g\prec_{0} g_{op}$ hold, then $g_{cl}$ and
$g_{op}$ are conformal, and thus, $g$ is isocausal to them.
\end{corollary}
{\it Proof.} From Proposition \ref{isostat'},
$\alpha(t)F^{\pm}(\cdot)\leqslant F^{\pm}(\cdot)$. Thus, $\alpha(t)\leqslant
1$ and:
\[
\int_{0}^{\infty}\frac{ds}{\replace{\sqrt{\alpha(s)}}{\alpha(s)}}\geq\int_{0}^{
\infty } 1\, ds=\infty , \qquad
\int_{-\infty}^{0}\frac{ds}{\replace{\sqrt{\alpha(s)}}{\alpha(s)}}\geq\int_{
-\infty}^{0}1\, ds=\infty;
\]
that is, condition
(\ref{intcond}) holds for $\tilde{\alpha}=\alpha$. Therefore, from
Proposition \ref{auxprop}, $g_{\ca}$ is conformal to $g_{\cc}$.
The last assertion is straightforward  from Remark \ref{rrr}. \cvd


\begin{remark}\label{remarksection4}
{\em Observe that all the constructions provided in Section
\ref{s3} ($B^\pm_{\tilde{\alpha}}(M),$
$M^+_{B,\tilde{\alpha}}\dots$) can be developed for both,
$\tilde{\alpha}=\alpha$ and $\tilde{\alpha}=1$. Moreover, when
$g_{\ca}$ and $g_{\cc}$ are conformal (in particular, under
condition (\ref{exp2})), such constructions are equivalent. In
these cases, we will denote such constructions with no mention to
$\tilde{\alpha}$.}
\end{remark}

\section{Partial boundaries $\hat\partial V, \check\partial V$ through isocausality}
\label{s5}

From now on, $V= (\R\times M,g)$ will denote a spacetime as in
(\ref{e1}) which satisfies condition (\ref{exp2}) for the metrics
in (\ref{exp1}) (unless other hypotheses are explicitly assumed),
so that Corollary \ref{cisocausal} is applicable. In particular,
the inequality $\alpha\leqslant 1$ holds (Proposition \ref{isostat'}).

\subsection{Information on $\hat \partial V$ at the point set level}

Along this section, we will use repeatedly the following set (see
(\ref{busemanfunc}) and (\ref{defbuse'})):
\begin{equation}\label{defbuse'''}
P_{\talpha}(b_{c,{\talpha}}^+) =\left\{(t_0,x_0)\in V:
\int_{0}^{t_0}
\frac{ds}{\replace{\sqrt{\talpha(s)}}{\talpha(s)}}<b_{c,{\talpha}}^+(x_0)
\right\}.
\end{equation}

\begin{convention}\label{conv1}
{\em The objects associated to the metric $g$ (resp. $g_{op}$;
$g_{cl}$) present no subindex (resp. subindex op; subindex cl). We
will focus on the future c-boundary $\hat
\partial V$ for $(V,g)$ (by comparing with $\hat \partial_{cl}
V\equiv \hat
\partial_{op} V$), being the arguments and results for the past
c-boundaries $\check\partial V, \check \partial_{cl} V\equiv
\check
\partial_{op} V$  totally
analogous (see Subsection \ref{j}).}
\end{convention}

As suggested in Remark \ref{rdis}, we are going to define a map
which takes $\hat{V}_{cl}$ into $\hat{V}$ by mapping every TIP
$P_{\cc}$ for $g_{cl}$ into the TIP $I^{-}(P_{cl})$ for $g$. In
order to establish under which conditions this map works, we need
to establish previously some technical results:
\begin{proposition}\label{p1}
If $\gamma:[a,\infty)\rightarrow V$, $\gamma(t)=(t,c(t))$ is a
future-directed timelike curve for $g$ such that $c(t)$ remains in
a bounded region of $(M,d^{+})$ then $I^{-}(\gamma)=V$.
\end{proposition}
{\it Proof.} Take any $(t_{0},x_{0})\in V$. From the hypothesis,
there exists $K>0$ such that $d^+(x_{0},c(t))<K$ for all $t\in
[a,\infty)$. In particular,
\begin{equation}\label{eq2}
d^+(x_{0},c(t))<K<\lim_{t\nearrow
\infty}\int_{t_{0}}^{t}ds=\infty.
\end{equation}
Therefore, from (\ref{eq2}) and Proposition \ref{p0} with
$\tilde{\alpha}\equiv 1$, one has $(t_{0},x_{0})\ll_{\cc}
(t,c(t))$ for $t$ big enough, thus $V=I^-_{\cc}(\gamma)\subset
I^-(\gamma)$ (the last inclusion because  $g_{cl}\prec_0 g$).
\cvd

\begin{proposition}\label{l11} Let $V$ be a spacetime as in
(\ref{e1}), with $g_{cl}$, $g_{op}$ as in (\ref{exp1}), and
satisfying (\ref{exp2}). Under the integral condition
\begin{equation}\label{e4}
\int_{0}^{\infty}\left(\frac{1}{\replace{\sqrt{\alpha(s)}}{\alpha(s)}}
-1\right)ds<\infty ,
\end{equation}
the following assertions hold:
\begin{itemize}
\item[(1)] Let $P_{\cc}=P_{\cc}(\bpcc{c})$ (recall
(\ref{defbuse'})) with $\bpcc{c}\in B_{\cc}^{+}(M)$ and
$c:[a,\Omega)\rightarrow M$, and put:
$$K_{\Omega}=\int_{0}^{\Omega}\left(\frac{1}{\replace{\sqrt{\alpha(s)}}{\alpha(s)}}
-1\right)ds.$$ Then, $\bpcc{c}+K_{\Omega}\in B_{op}^{+}(M)$ and
$I^-_{\ca}(P_{\cc})=P_{\ca}(\bpcc{c}+K_{\Omega})$. \item[(2)] If
$P^1_{\cc}\subsetneq P^2_{\cc}$  then
$I^-_{\ca}(P_{\cc}^1)\subsetneq I^-_{\ca}(P_{\cc}^2)$. Moreover,
if $P_{\cc}^1\neq P_{\cc}^2$ then
 $I^-_{\ca}(P_{\cc}^1)\neq
I^-_{\ca}(P_{\cc}^2)$.
\end{itemize}
\end{proposition}
{\it Proof.} (1) Consider the curve $\gamma(t)=(t,c(t))$, with
$t\in [a,\Omega)$, which is timelike for $g_{\cc}$ and satisfies
$I^-_{\cc}(\gamma)=P_{\cc}$. So,
$I^-_{\ca}(P_{\cc})=I^-_{\ca}(\gamma)=P_{\ca}(\bpca{c})$ where
$\bpca{c}$ is defined by (\ref{busemanfunc}) with $\talpha=\alpha$
(recall Convention \ref{conv1}).
But,
\[\begin{array}{rl} \bpca{c}(\cdot)= & \lim_{t\nearrow
\Omega}\left(\int_0^t(\frac{1}{\replace{\sqrt{\alpha(s)}}{\alpha(s)}}
-1)ds+t-d^+(\cdot , c(t))\right)\\ = & K_{\Omega}+\bpcc{c}(\cdot).
\end{array}\]
Hence, $I^{-}_{op}(P_{\cc})=P_{\ca}(\bpcc{c}+K_{\Omega})$ with
$\bpcc{c}+K_{\Omega}\in B^{+}_{op}(M)$ (notice that (\ref{e4})
implies $K_{\Omega}<\infty$).

\smallskip

(2) Consider $c_i:[a_i,\Omega_i)\rightarrow M$ such that
$P_{\cc}^i=P_{\cc}(\bpcc{c_i})$. From the inclusion
$P_{\cc}^1\subsetneq P_{\cc}^2$, we deduce that
$\bpcc{c_{1}}\leqslant \bpcc{c_{2}}$ with strict inequality at
some point. Moreover, $\Omega_1\leqslant \Omega_2$. In fact,
$d^+(c_1(s),c_1(t))<t-s$ for all $t,s\in [a_1,\Omega_1)$ with
$s<t$ and, then (recall that $t-d^+(p,c_i(t))$ is non-decreasing
with $t$),
$$s<t-d^+(c_1(s),c_1(t))\leqslant \lim_{t\nearrow
\Omega_1}\left(t-d^+(c_1(s),c_1(t))\right)=\bpcc{c_1}(c_1(s))\leqslant
\bpcc{c_2}(c_1(s))\leqslant \Omega_2,$$ for all $s\in [a_1,\Omega_1)$,
and thus, $\Omega_1\leqslant \Omega_2$ as required. Therefore,
$\bpcc{c_1}+K_{\Omega_1}\leqslant \bpcc{c_2}+K_{\Omega_2}$ with strict
inequality at some point. Taking into account that the left and
right members of the inequality characterize $P_{\ca}^1$ and
$P_{\ca}^2$, resp., it directly follows $P_{\ca}^1\subsetneq
P_{\ca}^2$.

\indent For the last assertion, assume by contradiction that
$P_{\cc}^1\neq P_{\cc}^2$ but
\begin{equation}\label{eqq1}I^-_{\ca}(P_{\cc}^1)=I^-_{\ca}(P_{\cc}^2).\end{equation}
From part (1), we deduce that
$K_{\Omega_1}+\bpcc{c_1}=K_{\Omega_2}+\bpcc{c_2}$. Suppose that
$K_{\Omega_2}\leqslant K_{\Omega_1}$ (the case $K_{\Omega_2}\geq
K_{\Omega_1}$ is analogous) and observe that, as
$\bpcc{c_2}=\bpcc{c_1} +(K_{\Omega_1}-K_{\Omega_2})$, necessarily
$P_{\cc}^1=P_{\cc}(\bpcc{c_1})\subset P_{\cc}(\bpcc{c_2})=
P_{\cc}^2$. From the first part of (2), we deduce
$I^{-}_{\ca}(P^1_{\cc})\varsubsetneq I^{-}_{\ca}(P^2_{\cc})$, in
contradiction to (\ref{eqq1}). \cvd

\begin{remark}\label{ex3bis} {\em In the proof of Proposition
\ref{l11}, condition (\ref{e4}) (namely, the causal cones of $g$
and $g_{\cc}$ approach at $t$-infinity) is only needed when the
IPs involved are of the form $P_{\cc}=P_{\cc}(\bpcc{c})$ with
$\Omega=\infty$, i.e. when $P_{\cc}\not\in \R\times M_{C}^{+}$
(recall (\ref{strucbor})). In fact, if $P_{\cc}\in \R\times M_C^+$
then the change $\bpcc{c}\mapsto \bpca{c} = \bpcc{c} + K_{\Omega}$
is just a re-scaling requested by (\ref{defbuse'''}). If
(\ref{e4}) does not hold then $K_{\Omega}=\infty$, and our
procedure takes every $P_{\cc}\not\in \R\times M_{C}^{+}$ into
$I_{\ca}^{-}(P_{\cc})\equiv V\equiv i^{+}$, being not one to one.

On the other hand, in the same proof condition (\ref{exp2}) is only needed to ensure that inequality $\alpha\leqslant 1$ holds.}
\end{remark}

\begin{theorem}\label{l1}
Let $V=(\R\times M,g)$ be a spacetime as in (\ref{e1}) such that
$g_{cl}\prec_0 g\prec_0 g_{op}$ holds. If (\ref{e4}) holds, then
the map
\[\begin{array}{cclc} \hat{j}:& \hat{V}_{\cc} & \rightarrow & \hat{V} \\
& P_{\cc} & \mapsto & I^-(P_{\cc})\end{array}\] is injective. So,
$\hat{V}$ contains a point set cone  of base $M_{B}^{+}$ and apex $i^+$.
\end{theorem}
{\it Proof.} 
Consider $P^1_{\cc}=P_{\cc}(\bpcc{c_1})$,
$P^2_{\cc}=P_{\cc}(\bpcc{c_2})$, with
$c_i:[a_i,\Omega_i)\rightarrow M$. Suppose that
$\hat{j}(P_{\cc}^1)=I^-(P_{\cc}^1)=I^-(P_{\cc}^2)=\hat{j}(P_{\cc}^2)$.
As $g\prec_0 g_{\ca}$, necessarily
$I^-_{\ca}(P_{\cc}^1)=I^-_{\ca}(P_{\cc}^2)$, and Proposition
\ref{l11} (2) ensures that $P_{\cc}^1=P_{\cc}^2$. For the last
assertion, recall that from (\ref{strucbor}) $\hat V_{\cc}\equiv
(\R\times M_{B}^{+})\cup\{i^{+}\}$. \cvd

Finally, we can particularize Theorem \ref{l1} to some relevant
cases:
\begin{corollary}\label{corollary1} Let $V=(\R\times M,g)$ be a spacetime as in
(\ref{e1}) such that
$g_{\cc}\prec_{0} g\prec_{0} g_{\ca}$.
\begin{itemize}
\item[(1)] If $(M,d^{+})$ is bounded, then $\partial_{\cal B}^+ M=\emptyset$ and $\hat{\partial} V$
contains a cone of base $\partial_C^+ M$ and apex $i^+$. If, in
addition, $(M,d^+)$ is complete (thus, compact), then
$\hat{\partial} V$ coincides with $i^+$. \item[(2)] If $(M,d^+)$
is complete and (\ref{e4}) holds, then $\partial_C^+ M=\emptyset$ and $\hat{\partial} V$ contains
a cone of base $\partial_{\cal B}^+ M$ and apex $i^+$ (recall
Remark \ref{remarksection4}).
\end{itemize}
\end{corollary}
{\it Proof.} (1) For the first assertion, as boundedness implies
$\partial_{\cal B}^+M=\emptyset$, the result follows from
Th. \ref{l1}. 
For the second one, note that any inextendible timelike curve
$\gamma:[a,\Omega)\rightarrow V$, $\gamma(t)=(t,c(t))$ must
satisfy $\Omega=\infty$: in fact, otherwise, Proposition \ref{p0}
implies ${\rm length}_{+}(c)<\infty$, and thus, $c(t)\rightarrow
x^+\in\partial_C^+ M$, in contradiction to the fact that
$\partial_C^+ M=\emptyset$. Therefore, from Proposition \ref{p1},
the past of any inextendible future-directed timelike curve
$\gamma$ is $V$.

(2) As completeness implies $\partial_C^+ M$ is empty, the result follows again by using Th. \ref{l1}. \cvd

\subsection{Discussion}

The following examples and comments show the optimality of the
hypotheses and results in previous subsection. The ambient
hypotheses $g_{\cc}\prec_{0} g\prec_{0} g_{\ca}$ will not be
discussed further, as it becomes  natural for isocausal comparison
(recall  Corollary \ref{cisocausal}) and imply $\alpha\leqslant 1$
from Proposition \ref{isostat'}. So, let us start with inequality
(\ref{e4}). As commented in Remark \ref{ex3bis}, this inequality
is necessary to ensure that $\hat j$ is one to one. This is
illustrated in the following trivial example.

\begin{example}\label{ex3} {\em Consider the following metrics on $V=\R\times
\R$:
\[
g_{\cc}=-dt^2+dx^2,\qquad g=-dt^2+dx^2/2,\qquad g_{\ca}=-dt^2+
dx^2/3.
\]
These metrics (all of them isometric) satisfy $g_{\cc}\prec_0
g\prec_0 g_{\ca}$, but condition (\ref{e4}) does not hold. Then,
any TIP $P_{\cc}$ for $g_{\cc}$, defined by a lightlike
curve\footnote{Here and in what follows, recall that  the
chronological past of any inextensible future-directed lightlike
curve is also a TIP (notice that the converse also holds, see
\cite[Sect 3.3]{FS}).} of the form $\gamma(t)=(t,\pm t)$,
satisfies $I^-(P_{\cc})=V$, being $\hat{j}(\hat{\partial}_{\cc}
V)=i^+$ (recall Remark \ref{ex3bis}).
}\end{example}

\begin{remark}\label{r1}{\em Even though condition (\ref{e4}) in Theorem
\ref{l1} solves the problem pointed out in Remark \ref{rdis}, the
following question arises now. From the definition of $\hat{j}$
and the relations $g_{\cc}\prec_0 g\prec_0 g_{\ca}$, the following
relations hold:
\[
P_{\cc}\subset \hat{j}(P_{\cc}),\quad
I^-_{\ca}(\hat{j}(P_{\cc}))=I^-_{\ca}(P_{\cc}).
\]
However, $\hat{j}(P_{\cc})$ may not be the unique TIP for $g$
satisfying these relations. As we will see next, this property is
closely related to the lack of surjectivity or continuity for
$\hat{j}$ in general.}
\end{remark}

The explicit example constructed in \cite{FHSnota} provides such a
non-surjective and non-continuous map\footnote{That example is
written in a different context, and does not satisfy (\ref{e4}).
However, it is trivial to modify it in order to ensure that
(\ref{e4}) is fulfilled. The example also has other remarkable
properties, as it shows explicitly a ``lightlike strain'' in
$\hat{\partial} V$ which forbides the existence of a isomorphism
(including the causal level) from $\hat{\partial}V$ to the
timelike c-boundary $\hat{\partial}_{\cc}V$.} $\hat j$. In that
example the contribution to the future c-boundary
$\hat{\partial}_{\cc}V$ comes completely from the forward Cauchy
boundary $\partial_C^+M$, i.e.,
$\hat{\partial}_{\cc}V=\left(\R\times
\partial_C^+M\right)\cup \{i^+\}$. Next, we are going to refine it to obtain
an example with $\partial_C^+M=\emptyset$, and so, ensuring that
only the proper Busemann boundary plays a role.

\begin{example}\label{e2}
(Even when $(M,d^{+})$ is complete -and (\ref{e4}) holds- $\hat j$
may be neither surjective nor continuous.) {\em Consider
$V=\R\times\R$ and three metrics $g_{\cc},g,g_{\ca}$ such that:
\begin{itemize}
 \item $g_{\cc}$ and $g_{\ca}$ are metrics as
 in (\ref{exp1}) with $\omega\equiv 0$, $h=dx^2$ and
$\replace{\alpha(t)=\frac{1}{(e^{-t}+1)^2}}{\alpha(t)=\frac{1}{e^{-t}+1}}$
(clearly, the inequality
 (\ref{e4})
 holds) and,
\item $g$ is a metric as in (\ref{e1}) with $\omega_{t}\equiv 0$.
\end{itemize}
In order to ensure condition (\ref{exp2}), we assume that $h_t$ satisfies
(recall Proposition \ref{isostat'}):
\begin{equation}\label{equejem}
 \alpha(t)dx^2\leqslant h_t\leqslant dx^2.
\end{equation}

Consider the curve $\gamma(t)=(t,t), t\in\R,$ which is lightlike
for $g_{\cc}$ and defines the TIP $P_{\cc}=I^-_{\cc}(\gamma)$.
Such a TIP is characterized by the Busemann function
$\bpcc{c}(\cdot)=\lim_{t\rightarrow
\infty}(t-d(\cdot,c(t)))<\infty$, with $c(t)=t$, and $d$ is the
Euclidean distance in $\R$. As $g_{\cc}\prec g_{\ca}$, $\gamma$ is
also causal for $g_{\ca}$, and so, it also defines a TIP for the
metric $g_{\ca},$ $P_{\ca}=I^-_{\ca}(\gamma)$. According to
expressions (\ref{defbuse}) and (\ref{defbuse'''}) with
$\tilde{\alpha}=\alpha$, this TIP is characterized by the
following function:
\[
\begin{array}{rl}
\bpca{c}(\cdot)= & \lim_{t\rightarrow \infty}\left( \int_{0}^t
\frac{1}{\replace{\sqrt{\alpha(s)}}{\alpha(s)}}ds-d(\cdot,t)\right) \\
 = & \lim_{t\rightarrow \infty}\left( \int_{0}^t
(\frac{1}{\replace{\sqrt{\alpha(s)}}{\alpha(s)}}-1)ds+(t-d(\cdot,t))\right)\\
= & \int_{0}^\infty e^{-s}ds + \lim_{t\rightarrow \infty}\left( t
-d(\cdot,t)\right)= 1+ b_{c,\cc}^+(\cdot)<\infty,
\end{array}
\]
i.e., $P_{\ca}\neq V$. Taking into account that $g_{\ca}$ is
(conformal to) a standard static spacetime, the result
\cite[Remark 4.17(1)]{FHSst} ensures the existence of some
reparametrization $\overline{c}$ of $c(t)=t$ such that
$\bpca{\overline{c}}=\bpca{c}$ and
$\alpha(t)|\dot{\overline{c}}(t)|=1$, i.e. the curve
$\bar\rho(t)=(t,\overline{c}(t))$, $t\in\R$ is lightlike for
$g_{\ca}$ (recall Remark \ref{remcharcau}).
Now, put $l=\overline{c}^{-1}$ and reparameterize $\bar \rho$ as
$\rho(s)=(l(s),s), s\in \R$. From $\bpca{\overline{c}}=\bpca{c}$,
it is deduced that $\lim_{s\rightarrow \infty}(l(s)-s)=0$. This
limit plus
$\dot{l}(s)=\replace{\sqrt{\alpha(l(s))}}{\alpha(l(s))}<1$ implies
$l(s)>s$ for all $s$ (see Figure \ref{f2}).

Let us assume now some additional hypotheses for $h_t$: suppose
that $h_t\equiv \alpha(t)dx^ 2$ (resp. $h_t\equiv dx^ 2$) at each
point $(t,x)$ such that $t\geq l(x)$ (resp. $t\leq x$), i.e.,
suppose that the metric $g$ coincides with $g_{\ca}$ above the
curve $\rho$ (resp. coincides with $g_{\cc}$ below the curve
$\gamma$).  Now, for each $k\in (0,l(0))$, define a function
$l_k:\R\rightarrow \R$ as the maximal solution of the following
initial value problem:

\begin{equation}\label{cauchyproblem}
 \left\{\begin{array}{l}
\dot{l}_{k}(t)=h_{l_{k}(t)}(\frac{\partial}{\partial x},\frac{\partial}{\partial
x}) \\
l_k(0)=k.
 \end{array}
\right.
\end{equation}

The curve $\gamma_{k}:\R\rightarrow \R^2$ defined by
$\gamma_{k}(t)=(l_k(t),t)$ satisfies:

\begin{itemize}
\item It is lightlike for $g$ (apply Remark \ref{remcharcau} with
$F^{+}_t=h_t$), thus it determines a TIP $P^k\in \hat{\partial}
V$.
\item $l_k(t)$ satisfies that $t< l_k(t)< l(t)$. Otherwise, as $0<k=l_k(0)<l(0)$, the curve $\gamma_k$ should intersect either $\rho$ or $\gamma$ at some point $t_0$ (see Figure \ref{f2}). But if, for example, $l_k(t_0)=l(t_0)$, the uniqueness of the solution to the first equality in (\ref{cauchyproblem}) with initial data at $t_0$ would imply $l_k\equiv l$ (recall that $h_t \equiv \alpha(t) dx^2$ at $\gamma_k(t_0) = \rho(t_0)$,
and the definitions of $l_k$ and $l$).
Therefore, $t< l_k(t)< l(t)$, and so,   $P_{\cc}\subset P^k\subset
P_{\ca}=I^-_{\ca}(P_{\cc})$. In particular, we deduce that $P_{\cc}\subset P^ k$ and $I^ -_{\ca}(P^ k)=P_{\ca}$.


\item By construction, if
$k_1,k_2\in (0,l(0))$ satisfy $k_1<k_2$, then $P^{k_1}\subsetneq
P^{k_2}$. In fact, the uniqueness of the solution for the initial
value problem ensures that $l_{k_1}(t)<l_{k_2}(t)$ for all $t$,
and so, the inclusion $\subset$ is trivially obtained. Moreover,
this inclusion is strict, since any past-directed timelike curve
from a point in $\gamma_{k_1}$ must remain below this curve (the
past cone at any point of $\gamma_{k_1}$ points out downward the
image of the curve).
\end{itemize}

Hence, for every $k\in (0,l(0))$ there exists a TIP $P^{k}$ for
$g$ such that $P_{\cc}\subset P^k\subset
I^{-}_{\ca}(P^{k})=I^-_{\ca}(P_{\cc})$. In particular, the map
$\hat{j}$ defined in Theorem \ref{l1} cannot be surjective since,
in that case, and according to Remark \ref{r1}, each $P^k$ should
be the image of $P_{\cc}$, which is impossible. Moreover,
$\hat{j}$ is not continuous; for instance, the sequence
$\{I^-_{\cc}(\gamma_{k}(n))\}_n$ converges to $P_{\cc}$ in
$(V,g_{\cc})$ but
$\{\hat{j}(I^-_{\cc}(\gamma_{k}(n)))\}_n=\{I^-(\gamma_{k}(n))\}_n$
converges to $P^k\neq \hat{j}(P_{\cc})=I^-(P_{\cc})$ in $(V,g)$.

\begin{figure}
\centering
\ifpdf
  \setlength{\unitlength}{1bp}%
  \begin{picture}(211.21, 154.42)(0,0)
  \put(0,0){\includegraphics{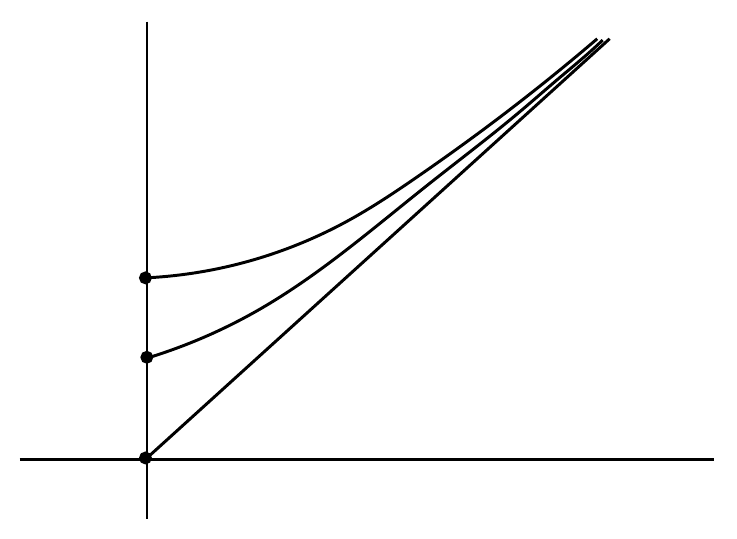}}
  \put(28.47,23.55){\fontsize{8.54}{10.24}\selectfont $0$}
  \put(23.11,73.63){\fontsize{8.54}{10.24}\selectfont $l(0)$}
  \put(70.50,42.33){\fontsize{8.54}{10.24}\selectfont $\gamma$}
  \put(56.19,81.23){\fontsize{8.54}{10.24}\selectfont $\rho$}
  \put(50.79,60.15){\fontsize{8.54}{10.24}\selectfont $\gamma_k$}
  \put(29.77,49.42){\fontsize{8.54}{10.24}\selectfont $k$}
  \end{picture}%
\else
  \setlength{\unitlength}{1bp}%
  \begin{picture}(211.21, 154.42)(0,0)
  \put(0,0){\includegraphics{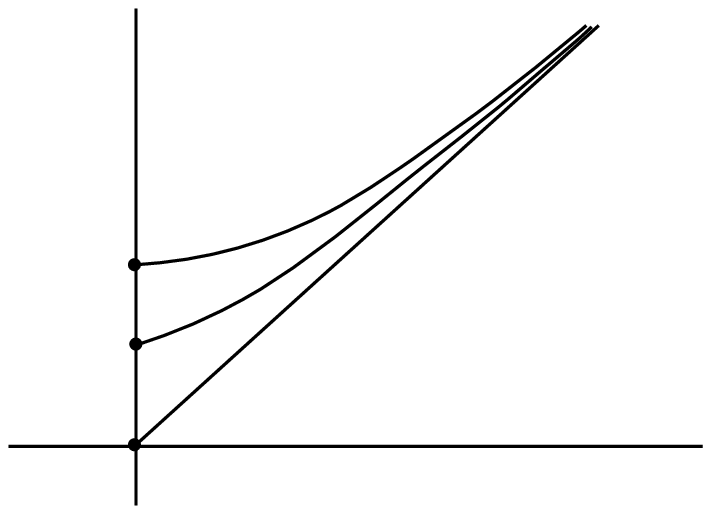}}
  \put(28.47,23.55){\fontsize{8.54}{10.24}\selectfont $0$}
  \put(23.11,73.63){\fontsize{8.54}{10.24}\selectfont $l(0)$}
  \put(70.50,42.33){\fontsize{8.54}{10.24}\selectfont $\gamma$}
  \put(56.19,81.23){\fontsize{8.54}{10.24}\selectfont $\rho$}
  \put(50.79,60.15){\fontsize{8.54}{10.24}\selectfont $\gamma_k$}
  \put(29.77,49.42){\fontsize{8.54}{10.24}\selectfont $k$}
  \end{picture}%
\fi \caption{\label{f2} {\it Qualitative behavior of functions $\rho$ ,$\gamma_{n}$ and $\gamma$ in Example \ref{e2}.}}
\end{figure}
}
\end{example}
\subsection{Strains and topological information}

A reason for the lack of continuity of $\hat{j}$ is the existence of different TIPs for the metric $g$
sharing the same past when computed with the metric $g_{\ca}$
(recall Remark \ref{r1}). In principle, this also indicates the existence of
alternative choices for $\hat{j}$. In order to overcome this
problem, we are going to identify all these TIPs by passing to the
corresponding quotient space. To this aim, first we need the
following technical lemma.
\begin{lemma}\label{lemmasobre}
 Suppose that (\ref{e4}) holds and consider an IP $P\in \hat{V}$ for the
metric $g$ and $P_{\ca}:=I^{-}_{\ca}(P)$. Then, there exists an unique 
$P_{\cc}\in \hat{V}_{\cc}$ satisfying:
\[
 P_{\cc}\subset P \;\;\hbox{and}\;\; I^-_{\ca}(P_{\cc})=P_{\ca}.
\]
\end{lemma}
{\it Proof. } Suppose that $P=I^{-}(\gamma)$ with
$\gamma:[a,\Omega)\rightarrow V$, $\gamma(s)=(s,c(s))$ a
future-directed timelike curve for $g$. Observe that, assumed the existence of such a TIP $P_{\cc}$, the uniqueness follows from Proposition \ref{l11} (2). For the existence, we will consider two cases for clarity.

In one hand, assume that $\Omega<\infty$. In this case, $c(s)\rightarrow x^+\in \partial_C^+ M$ as $s\rightarrow \Omega$
($\gamma$ is causal in $(V,g_{\ca})$). Consider a timelike curve
for $g_{\cc}$, $\gamma_{\cc}:[a,\Omega)\rightarrow V$, of the form
$\gamma_{\cc}(s)=(t_{\cc}(s),c(s))$ where $t_{\cc}(s)<s$ and
$t_{\cc}(s)\rightarrow \Omega$. This curve can be constructed by
taking
\[\dot{t}_{\cc}(s)>\frac{1}{\replace{\sqrt{\alpha_0}}{\alpha_0}}>
\frac{F^{+}_s(\dot{c}(s))}{\replace{\sqrt{\alpha_0}}{\alpha_0}}\geq
F^{+}(\dot{c}),\] where $\alpha_0={\rm
min}|_{[a,\Omega]}(\alpha(s))$ (recall that $\gamma$ is timelike
for $g$, i.e. $F^+_s(\dot{c})<1$, and (\ref{isostat})). From the
construction, $P_{\cc}=I^-_{\cc}(\gamma_{\cc})\subset P$ and, as
both curves end at the same pair $(\Omega,x^+)$, necessarily
$I^-_{\ca}(P_{\ca})=I^-_{\ca}(P)$ (recall (\ref{ddd})).
%
%

In the other hand, suppose that $\Omega=\infty$. First, consider the curve
$\gamma_{cl}(s)=(t_{cl}(s),c(s))$, with $s\in [a,\infty)$ and where the component $t_{cl}(s)$
is determined by the Finsler metrics in (\ref{charcau}) and
(\ref{charcau'}) as:
\[
\frac{dt_{cl}}{ds}=\frac{F^{+}(\dot{c})}{F^{+}_{s}(\dot{c})},\qquad
t_{cl}(a)=-k_{cl}:=-\int_{a}^{\infty}\left(\frac{F^{+}(\dot{c})}{F^{+}_{s}(\dot{
c })}-1\right).
\]
Note that $k_{cl}$ is finite. In fact, from (\ref{isostat}) it is
$1\leqslant F^{+}(\dot{c})/F^{+}_{s}(\dot{c})\leqslant
1/\replace{\sqrt{\alpha(s)}}{\alpha(s)}$,
which joined to (\ref{e4}) imply
\[
(0\leqslant) k_{cl}=\int_{a}^{\infty}\left(\frac{F^{+}(\dot{c})}{F^{+}_{s}(\dot{c})}
-1\right)\leqslant\int_{a}^{\infty}\left(\frac{1}{\replace{\sqrt{\alpha(s)}}{\alpha(s)}}
-1\right)<\infty.
\]
Moreover, as the curve $\gamma(s)$ is future-directed timelike for
$g$, i.e. $F^{+}_{s}(\dot{c})<1$, so is the curve $\gamma_{cl}$
for $g_{cl}$, since
\[
F^{+}(\dot{c})<\frac{F^{+}(\dot{c})}{F^{+}_{s}(\dot{c})}=\frac{dt_{cl}}{ds}.
\]
Similarly, one can consider the curve
$\gamma_{op}(s)=(t_{op}(s),c(s))$, where the component $t_{op}(s)$
is determined by:
\[
\frac{dt_{op}}{ds}=\frac{\replace{\sqrt{\alpha(s)}}{\alpha(s)}F^{+}(\dot{c})}{F^
{ + } _ { s } (\dot { c } ) } , \qquad
t_{op}(a)=k_{op}:=\int_{a}^{\infty}\left(1-\frac{\replace{\sqrt{\alpha(s)}}{
\alpha(s)}F^{+}(\dot{c})}
{ F^{+}_{s}(\dot{c})}\right).
\]
Again, $k_{op}$ is finite since, from (\ref{isostat}), it is
$\replace{\sqrt{\alpha(s)}}{\alpha(s)}\leqslant
\replace{\sqrt{\alpha(s)}}{\alpha(s)}F^{+}(\dot{c})/F^{+}_{s}(\dot{c})\leqslant
1$, which joined to (\ref{e4}) implies
\[\begin{array}{rl}
(0\leqslant)k_{op}
&=
\int_{a}^{\infty}\left(1-\frac{\replace{\sqrt{\alpha(s)}}{\alpha(s)}F^{+}(\dot{c
} ) } { F^ {
+ } _ { s }
(\dot{c})}\right)\leqslant\int_{a}^{\infty}(1-\replace{\sqrt{\alpha(s)}}{\alpha(s)}
)=\\
&= \int_{a}^{\infty}
\replace{\sqrt{\alpha(s)}}{\alpha(s)}\left(\frac{1}{\replace{\sqrt{\alpha(s)}}{
\alpha(s)} } -1\right)\leqslant\int_ { a } ^ { \infty }
\left(\frac{1}{\replace{\sqrt{\alpha(s)}}{\alpha(s)}}-1\right)<\infty.
\end{array}\]
With this definition, the curve $\gamma_{\ca}$ is future-directed
timelike for $g_{\ca}$, since
\[
\replace{\sqrt{\alpha(s)}}{\alpha(s)}F^{+}(\dot{c})<\replace{\sqrt{\alpha(s)}}{
\alpha(s)} \frac { F^ { + } (\dot { c } ) } { F^ { + } _ { s }
(\dot{c})}=\frac{dt_{op}}{ds}.
\]
Note also that $\lim_{s\rightarrow\infty}(t_{op}(s)-t_{cl}(s))=0$.
In fact:
\begin{equation}\label{y}
\begin{array}{rl}
\lim_{s\rightarrow\infty}(t_{op}(s)-t_{cl}(s)) &
=\lim_{s\rightarrow\infty}(t_{op}(s)-s+s-t_{cl}(s)) \\ &
=\lim_{s\rightarrow\infty}\left(\int_{a}^{s}\frac{dt_{op}}{ds'}ds'+k_{op}
-s+s-\int_{a}^{s}\frac{dt_{cl}}{ds'}ds'+k_{cl}\right)
\\ &
=\lim_{s\rightarrow\infty}\left(\int_{a}^{s}\left(\frac{dt_{op}}{ds'}
-1\right)ds'+k_{op}-\int_{a}^{s}\left(\frac{dt_{cl}}{ds'}-1\right)ds'+k_{cl}
\right)
\\ &
=\int_{a}^{\infty}\left(\frac{dt_{op}}{ds}-1\right)ds-\int_{a}^{\infty}\left(\frac
{ dt_{cl}}{ds}-1\right)ds+k_{op}+k_{cl}
\\ &
=\int_{a}^{\infty}\left(\frac{\replace{\sqrt{\alpha(s)}}{\alpha(s)}F^{+}(\dot{c})}{F^{+}_{s}(\dot{c})}-1\right)ds-\int_{a}^{\infty}\left(\frac{F^{+}(\dot{c})}{F^{+}_{s}(\dot{c})}
-1\right)ds+k_{op}+k_{cl}
\\ & =0.
\end{array}
\end{equation}
Hence, by construction it is clear that $t_{cl}(s)\leqslant s\leqslant
t_{op}(s)$ for all $s\in [a,\infty)$. Therefore, if one defines
the following TIPs for $g_{cl}$ and $g_{op}$, resp.,
\[
P_{cl}:=I^{-}_{cl}(\gamma_{cl})\qquad
P^*_{op}:=I^{-}_{op}(\gamma_{op}),
\]
one automatically has these inclusions:
\[
P_{cl}\subset P\subset P^*_{op},\quad\hbox{and thus}\quad
I^{-}_{op}(P_{cl})\subset P_{\ca}(=I^-_{\ca}(P))\subset
I^{-}_{op}(P^*_{op})=P^*_{op}.
\]
So, in order to obtain the required equality $I^-_{\ca}(P_{\cc})=P_{\ca}$, it is sufficient to prove
\begin{equation}\label{t}
I^{-}_{op}(P_{cl})=P^*_{op}.
\end{equation}
Define the curves $c_{cl}(s):=c(t_{cl}^{-1}(s))$,
$c_{op}(s):=c(t_{op}^{-1}(s))$. Then,
\[
\begin{array}{rl}
P_{cl}=I^{-}_{cl}(\tilde{\gamma}_{cl}), & \qquad
\hbox{where}\;\;\tilde{\gamma}_{cl}(s):=(s,c_{cl}(s))
\\
P^*_{op}=I^{-}_{op}(\tilde{\gamma}_{op}), & \qquad
\hbox{where}\;\;\tilde{\gamma}_{op}(s):=(s,c_{op}(s))
\end{array}
\]
From the study developed in Section \ref{ff}, the TIPs
$I^{-}_{op}(P_{cl})$, $P^*_{op}$ can be rewritten as
follows (recall (\ref{defbuse}) and (\ref{busemanfunc})):
\[
%
%
I^{-}_{op}(P_{cl})=P_{op}(\bpca{c_{cl}}),\qquad
P^*_{op}=P_{op}(\bpca{c_{op}})
\]
So, (\ref{t}) will follow if we prove that
$\bpca{c_{op}}=\bpca{c_{cl}}$. To this aim, first
note that
\[
\begin{array}{l}
\bpca{c_{cl}}(\cdot)=\lim_{s\rightarrow\infty}\left(\int_{0}^{s}
\frac {
ds'}{\replace{\sqrt{\alpha(s')}}{\alpha(s')}}-d^{+}(\cdot,c_{cl}
(s))\right)=\lim_ { s\rightarrow\infty }
\left(\int_{0}^{t_{cl}(s)}\frac{ds'}{\replace{\sqrt{\alpha(s')}}{\alpha(s')}}-d^
{ + } (\cdot , c(s))\right)
\\
\bpca{c_{op}}(\cdot)=\lim_{s\rightarrow\infty}\left(\int_{0}^{s}
\frac {
ds'}{\replace{\sqrt{\alpha(s')}}{\alpha(s')}}-d^{+}(\cdot,c_{op}
(s))\right)=\lim_ { s\rightarrow\infty }
\left(\int_{0}^{t_{op}(s)}\frac{ds'}{\replace{\sqrt{\alpha(s')}}{\alpha(s')}}-d^
{ + } (\cdot , c(s))\right).
\end{array}
\]
Thus, from (\ref{e4}) and (\ref{y}), one deduces
\[
\begin{array}{rl}
\bpca{c_{op}}(\cdot)-\bpca{c_{cl}}(\cdot) &
=\lim_{s\rightarrow\infty}(\int_{0}^{t_{op}(s)}\frac{ds'}{\replace{\sqrt{
\alpha(s') }}{\alpha(s')} }
-\int_{0}^{t_{cl}(s)}\frac{ds'}{\replace{\sqrt{\alpha(s')}}{\alpha(s')}})
\\ &
=\lim_{s\rightarrow\infty}(\int_{t_{cl}(s)}^{t_{op}(s)}\frac{ds'}{\replace{\sqrt
{ \alpha(s')}}{\alpha(s')}})
\\ &
=\lim_{s\rightarrow\infty}\left(\int_{t_{cl}(s)}^{t_{op}(s)}\left(\frac{1}{
\replace{\sqrt
{ \alpha(s')}}{\alpha(s')}}-1\right) ds' + \int_{t_{cl}(s)}^{t_{op}(s)}
ds'\right)
\\ & =0.
\end{array}
\]

\cvd

\smallskip

We are now in conditions to introduce the following definition.

\begin{definition}\label{strain} Two  TIPs $P^1,P^2\in\hat{\partial} V$ are {\em st-related},
$P^1\sim_{st} P^2$, if there exists a TIP $P_{\cc}$ for $g_{\cc}$
such that $P_{\cc}\subset P^1 \cap P^2$ and
$I^-_{\ca}(P^1)=I^-_{\ca}(P_{\cc})=I^-_{\ca}(P^2)$.

When (\ref{e4}) holds, this relation is of
equivalence\footnote{Symmetry and transitivity are
straightforward, while reflexivity is a consequence of Lemma
\ref{lemmasobre}.} and it defines the quotient space
$\hat{V}/\sim_{st}$. The (non-trivial) classes of equivalence in
$\hat{V}/\sim_{st}$ will be called {\em (future)
strains}.

We will denote by $\hat{\Pi}$ the natural projection onto the
quotient, $\hat{\Pi}: \hat{V} \rightarrow \hat{V}/\sim_{st}$.
\end{definition}

\begin{remark}{\em
The space $\hat{V}/\sim_{st}$ is endowed with the quotient
topology from $\hat{V}$. Here, we are going to justify that this
quotient topology can be also given as well in terms of a limit
operator. The argument below is general.

Let $L$ be a limit operator on a set $X$, i.e., a map
$L:S(X)\rightarrow {\cal P}(X)$ (where $S(X)$ is the set of all
sequences in $X$ and ${\cal P}(X)$ is the set of parts of $X$)
satisfying that, if $\tilde{\sigma}$ is a subsequence of some
$\sigma\in S(X)$, then $L(\sigma)\subset L(\tilde{\sigma})$. Endow
$X$ with the associated topology $\tau^{L}$, defined by the
following statement: $C\subset X$ is {\em closed} if, and only if,
$L(\sigma)\subset C$ for all sequence $\sigma$ in $C$. Define in
$X$ a relation of equivalence $\sim$ and denote by
$\pi:X\rightarrow X/\sim$ the natural projection. The space
$X/\sim$ is endowed with the quotient topology induced from
$\tau^{L}$ and $\sim$. Alternatively, note that $L$ and $\sim$
allow to define a limit operator $\LQG$ in the quotient $X/\sim$
just by putting:\footnote{It is worth pointing out that this is
not the unique possible choice for $\LQG$ (a drawback of $\LQG$ is
that, even if $L$ is of the first order, $\LQG$ may not). However,
other natural alternatives would yield the same topology.}
\begin{equation}\label{deflimitQ}[x]\in \LQG (\{[x_n]\})\iff \exists\;\;
x'\in \pi^{-1}([x]),\; x'_n\in \pi^{-1}([x_n])\;\,\forall n\in\N\;
:\,\; x'\in L(\{x'_n\}).\end{equation} where $[x],[x_n]\in
X/\sim$. The equivalence between both topologies on $X/\sim$, the
topology $\tau^{\LQG}$ associated to $\LQG$ and the quotient
topology $\tau^{L}/\sim$ induced from $\tau^{L}$ and $\sim$, comes
from the following equivalences:
\[
\begin{array}{rl} C\subset X/\sim \hbox{ is closed for } \tau/\sim
\iff & \pi^{-1}(C) \hbox{ is closed in }
X \\  \iff & L(\sigma)\subset \pi^{-1}(C) \hbox{ for all
sequence } \sigma\subset \pi^{-1}(C)\\ \iff &
\LQG(\pi(\sigma))\subset C \hbox{ for all sequence }\pi(\sigma)\subset
C.
\end{array}
\]
}\end{remark}

\smallskip

Next, we are going to determine when the composition
$\apf:=\hat{\Pi}\circ\hat{j}$ is bijective and continuous. But,
first, we need some previous lemmas.

\begin{lemma}\label{l111} Assume that (\ref{e4}) holds, $d_{Q}^+$ is a generalized distance and
$M_{C}^+$ is locally compact. Then, $P_{\cc}\in
\hat{L}(\{P_{\cc}^n\})$ if and only if $P_{\ca}\in
\hat{L}(\{P^{n}_{\ca}\})$, where $P_{\ca}=I^-_{\ca}(P_{\cc})$ and
$P_{\ca}^n=I^-_{\ca}(P_{\cc}^n)$.
\end{lemma}
{\it Proof.} Consider $\gamma_{n}:[a_n,\Omega_n)\rightarrow V$,
$\gamma:[a,\Omega)\rightarrow V$ such that
$\gamma_{n}=(t,c_n(t))$, $\gamma(t)=(t,c(t))$ and
$P_{\cc}^n=P_{\cc}(\bpcc{c_n})=I^-_{\cc}(\gamma_n)$,
$P_{\cc}=P_{\cc}(\bpcc{c})=I^-_{\cc}(\gamma)$. Recalling
(\ref{defbuse}), (\ref{busemanfunc}), we deduce that
$I^-_{\ca}(\gamma_n)=P_{\ca}(\bpca{c_n})$ and
$I^-_{\ca}(\gamma)=P_{\ca}(\bpca{c})$. In particular, our
statement is equivalent to the following one (recall Section
\ref{s3.3}): \[\bpcc{c}\in \hat{L}(\{\bpcc{c_n}\})\; \hbox{if and
only if}\; \bpca{c}\in\hat{L}(\{\bpca{c_n}\}).\]
We will focus on the implication to the left, as the other one is
similar but simpler. So, assume that
$\bpca{c}\in\hat{L}(\{\bpca{c_n}\})$. From Proposition \ref{l11},
we have that $\bpca{c_n}=\bpcc{c_n} + K_{\Omega_n}$ and
$\bpca{c}=\bpcc{c} + K_{\Omega}$. Then, from our assumption, we
deduce that $\bpcc{c}\in \hat{L}\left(\{\bpcc{c_n} +
\left(K_{\Omega_n}-K_{\Omega}\right)\}_{n}\right)$. The result
follows if we prove that $K_{\Omega_n}\rightarrow K_{\Omega}$ or,
equivalently, that $\Omega_n\rightarrow \Omega$ (recall
(\ref{e4})).

On one hand, suppose that $\Omega<\infty$ and define:
\begin{equation}\label{omegaop}
 \Omega_{\ca}^n=\int_0^{\Omega_n}\frac{ds}{\replace{\sqrt{\alpha(s)}}{\alpha(s)}},\qquad
\Omega_{\ca}=\int_0^{\Omega}\frac{ds}{\replace{\sqrt{\alpha(s)}}{\alpha(s)}}
\end{equation}
With these definitions, we obtain that
$\bpca{c}=d^+_{(\Omega_{\ca},x^+)}$ for some $x^+\in M_{C}^+$
(recall (\ref{ddd2}) and (\ref{ddd1})). As $M_C^+$ is locally
compact, $d_Q^+$ is a generalized distance and
$d^+_{(\Omega_{\ca},x^+)}=\bpca{c}\in \hat{L}(\{\bpca{c_n}\})$,
the naturally adapted version of \cite[Proposition 5.24]{FHSst}
ensures that $\{\Omega^n_{\ca}\}\rightarrow \Omega_{\ca}$, and so,
that $\Omega_n\rightarrow \Omega$.

On the other hand, if $\Omega=\infty$, as $\gamma(t)\in P_{\ca}\subset
\LI(\{P_{\ca}^n\})$, for every $t\in [a,\infty)$ there exists $n_0$
such that $\gamma(t)=(t,c(t))\in P_{\ca}^n$ for $n\geq n_0$.
Therefore, $\gamma(t)\ll_{\ca} \gamma_n(t')$ for some $t'\in
[a_n,\Omega_n)$, and, from Prop. \ref{p0},
$$\int^{t'}_{t}\frac{ds}{\replace{\sqrt{\alpha(s)}}{\alpha(s)}}>d^+(c(t),
c_n(t'))\geq
0,$$ which implies that $\Omega_{n}(>t')\geq t$. Taking $t\nearrow
\infty$, one deduces that $\Omega_{n}\rightarrow \Omega=\infty$. \cvd

\begin{lemma}\label{propapf}
Under the hypotheses of Lemma \ref{l111}, if $P_{\cc}\in
\hat{L}(\{P_{\cc}^n\})$ then any sequence $\{P^n\}_{n}$, with
$P^n\in \hat{\Pi}^{-1}(\apf (P_{\cc}^n))$, contains some
subsequence $\{P^{n_k}\}_{k}$ such that $P\in
\hat{L}(\{P^{n_k}\})$  for some $P\in \hat{\Pi}^{-1}(\apf
(P_{\cc}))$.
\end{lemma}
Before the proof of this lemma, we have to introduce the following
general result for c-completions (recall \cite[Corollary
5.12]{FH}). For the convenience of the reader, a detailed proof of
this result is included here.
\begin{lemma}\label{auxiliar}
Take $\{P^n\}\subset \hat{V}$ and $P'\in \hat{V}$. If $P'\subset
\LI(\{P^n\})$ then there exist $P\supset P'$ and a subsequence
$\{P^{n_k}\}_k\subset \{P^n\}$ such that $P\in
\hat{L}(\{P^{n_k}\}_k)$.
\end{lemma}
{\it Proof. } As $P'\subset \LI(\{P^n\})\subset \LS(\{P^n\})$,
Zorn's lemma ensures the existence of a maximal IP $P$ such that
$P'\subset P\subset \LS(\{P^n\})$. Let $\{x_m\}\subset V$ be a
chronological chain (i.e. $x_{m}\ll x_{m+1}$ for all $m$)
generating $P$ (i.e. $P=\cup_m I^-(x_m)$). As $P\subset
\LS(\{P^n\})$, for every $m$ there exists
$\{n^{m}_{k}\}_{k}\subset \N$ such that $x_m\in P^{n^{m}_{k}}$ for
all $k$. As $x_{m}\ll x_{m+1}$, the
desired subsequence is $\{P^{n_{k}}\}_{k}$ with
$n_{k}:=n^{k}_{k}$. In fact, by construction $P$ satisfies: (i)
$P\subset \LI(\{P^{n_k}\})$, (ii) $P$ is maximal in
$\LS(\{P^{n_k}\})$ (as $P$ is maximal in $\LS(\{P^n\})$), and the
first assertion follows. \cvd
%

{\it Proof of Lemma \ref{propapf}.} As $P_{\cc}\in
\hat{L}(\{P_{\cc}^n\})$, necessarily $P_{\cc}\subset
\LI(\{P^{n}_{\cc}\})$, and thus, $P_{\cc}\subset
{\LI}(\{I^{-}(P^{n}_{\cc})\})$. As the inferior limit is a past
set for $g$, necessarily
\begin{equation}\label{ju}\hat{j}(P_{\cc})=I^-(P_{\cc})\subset
\LI(\{I^-(P_{\cc} ^n)\})=\LI(\{\hat{j}(P^n_{\cc})\}).\end{equation}
Moreover, since any element $P^ n\in
\hat{\Pi}^{-1}(\apf(P_{\cc}^n))$ satisfies $P^n_{\cc}\subset P^n$
(Defn. \ref{strain}), necessarily $I^{-}(P^{n}_{\cc})\subset
I^{-}(P^{n})=P^{n}$. Taking into account (\ref{ju}), it is
$$\hat{j}(P_{\cc})\subset\LI(\{I^-(P_{\cc}^n)\}) \subset
\LI(\{P^n\})\subset \LS(\{P^n\}),$$ and
$\hat{j}(P_{\cc})$ is an IP for $g$. Then, from Lemma
\ref{auxiliar} there exist an IP $P$ for $g$ and some subsequence
$\{P^{n_k}\}_{k}\subset\{P^{n}$\} such that:
\begin{equation}\label{eq3}P\in \hat{L}(\{P^{n_k}\}).\end{equation}
In order to check that $P\in \hat{\Pi}^{-1}(\apf (P_{\cc}))$, note
that from $g\prec_0 g_{\ca}$ and the inclusion
$\hat{j}(P_{\cc})\subset P$,
\begin{equation}\label{ecuaux2}
I^{-}_{\ca}(P_{\cc})=I^{-}_{\ca}(I^{-}(P_{\cc}))\subset
I^-_{\ca}(P).
\end{equation}
Taking into account that $P^{n_{k}}\in\hat{\Pi}^{-1}(\hat{{\cal
J}}(P^{n_{k}}_{cl}))$ and (\ref{eq3}), it is straightforward that
\begin{equation}\label{ecuaux} I^-_{\ca}(P)\subset
\LI(\{I^-_{\ca}(P^{n_k})\}))=\LI(\{I^-_{\ca}(P^{n_k}_{\cc})\}).\end{equation}
Moreover, as $P_{\cc}\in\hat{L}(\{P^n_{\cc}\})$, from Lemma
\ref{l111} the subsequence satisfies $I^-_{\ca}(P_{\cc})\in
\hat{L}(\{I^-_{\ca}(P^{n_k}_{\cc})\})$. Hence,
$I^-_{\ca}(P_{\cc})$ is maximal in
$LS(\{I^-_{\ca}(P^{n_k}_{\cc})\})$ (recall (\ref{limcrono})),
which joined to (\ref{ecuaux2}) and (\ref{ecuaux}) implies
\[
I^-_{\ca}(P_{\cc})= I^-_{\ca}(P).
\]
From this equality and the fact that $P_{\cc}\subset
I^-(P_{\cc})\subset P$, we deduce that $P\in \hat{\Pi}^{-1}(\apf
(P_{\cc}))$, as required.\cvd

%
%
%

\begin{theorem}\label{apfc}
Let $V=(\R\times M,g)$ be a spacetime as in (\ref{e1}) such that
$g_{\cc}\prec_0 g\prec_0 g_{\ca}$. If (\ref{e4}) holds, $d_Q^+$ is
a generalized distance and $M_C^+$ is locally compact then the map
$$\apf=\hat{\Pi}\circ\hat{j}:\hat{V}_{\cc} \rightarrow \hat{V}/\sim_{st}$$ is bijective and
continuous. Moreover, if, in addition, $\hat{V}/\sim_{st}$ is
Hausdorff, then $\apf$ is an homeomorphism.
\end{theorem}
{\it Proof.} For the injectivity of $\apf$, assume that
$P^1_{\cc}\neq P^2_{\cc}$. From Proposition \ref{l11} (2),
$I^-_{\ca}\left(\hat{j}(P^1_{\cc})\right)=I^-_{\ca}(P_{\cc}^1)\neq
I^-_{\ca}(P_{\cc}^2)=I^-_{\ca}\left(\hat{j}(P_{\cc}^2)\right)$. So,
$\hat{j}(P_{\cc}^1)\not\sim_{st} \hat{j}(P_{\cc}^2)$ and then
$\apf(P_{\cc}^1)\neq \apf (P_{\cc}^2)$.

For the continuity\footnote{\label{foot1} Notice that the simple
characterization of continuity by means of sequences (which is
used typically in first countable spaces) holds also when the
domain is a sequential space defined by a limit operator, see
\cite[Footnote 13]{FHSconf}.} of $\apf$, consider a sequence
$\{P_{\cc}^n\}\subset \hat{V}_{\cc}$ and $P_{\cc}\in
\hat{V}_{\cc}$ such that $P_{\cc}\in \hat{L}(\{P_{\cc}^n\})$. By
contradiction, suppose that $\{\apf(P_{\cc}^n)\}$ does not
converge to $\apf(P_{\cc})$. Then, there exists an open
neighborhood $U$ of $\apf(P_{\cc})$ such that, up to a
subsequence, it does not contain $\{\apf(P_{\cc}^{n})\}$ for any
$n$; i.e., $\hat{\Pi}^{-1}(U)$ is an open set of $\hat{V}$
containing all the elements in $\hat{\Pi}^{-1}(\apf(P_{\cc}))$,
but not containing any element of
$\hat{\Pi}^{-1}(\apf(P_{\cc}^{n}))$. This is in contradiction to
the fact that, from Lemma \ref{propapf}, there exists some
subsequence $\{P^{n_{k}}\}_{k}$, $P^{n_{k}}\in
\hat{\Pi}^{-1}(\hat{{\cal J}}(P^{n_{k}}_{cl}))$ and $P\in
\hat{\Pi}^{-1}(\hat{{\cal J}}(P_{cl}))$ such that $P\in
\hat{L}(\{P^{n_k}\})$.

For the surjectivity of $\apf$, consider an arbitrary element
$\hat{\Pi}(P)\in \hat{V}/\sim_{st}$, with $P$ IP of $(V,g)$. Lemma
\ref{lemmasobre} ensures the existence of $P_{\cc}\in
\hat{V}_{\cc}$ such that
\[
                                           P_{\cc}\subset P \;\; \hbox{and}\;\;
I^-_{\ca}(P_{\cc})=I^-_{\ca}(P).
                                          \]
In particular, we deduce that $(\hat{j}(P_{\cc})=)I^-(P_{\cc})\sim_{st} P$, and
so, that $\apf(P_{\cc})=\hat{\Pi}(\hat{j}(P_{\cc}))=\hat{\Pi}(P)$, as required.

Finally, let us prove that $\hat{{\cal J}}^{-1}$ is continuous if
$\hat{V}/\sim_{st}$ is Hausdorff. Consider a sequence
$\{\hat{{\cal J}}(P^{n}_{\cc})\}_n\subset \hat{V}/\sim_{st}$ and
$\hat{{\cal J}}(P_{\cc})\in\hat{V}/\sim_{st}$ such that
$\hat{{\cal J}}(P_{\cc})\in \LQ(\{\hat{{\cal J}}(P^{n}_{\cc})\})$
(where $\LQ$ is the quotient limit associated to $\hat{L}$ and
$\sim_{st}$, recall (\ref{deflimitQ})). We have just to check that
$P_{\cc}\in \hat{L}(\{P^{n}_{cl}\})$, and so, $\{P_{\cc}^n\}$
converges to $P_{\cc}$ with the future chronological topology
(recall again footnote \ref{foot1}). Assume by contradiction that this
is not the case. From the definition of $\LQ$, there exist a sequence
$\{P^n\}$, with $P^n\in \hat{\Pi}^{-1}(\hat{{\cal
J}}(\{P^{n}_{\cc}\}))$, and some $P\in \hat{\Pi}^{-1}(\hat{{\cal
J}}(P_{\cc}))$ such that $P\in \hat{L}(\{P^n\})$. In particular,
from (\ref{limcrono}) and the fact that $g\prec g_{\ca}$,
$P_{\ca}:=I^-_{\ca}(P)\subset
LI(\{I^-_{\ca}(P^n)\})=LI(\{P_{\ca}^n\})$. Observe that
$P_{\ca}\notin \hat{L}(\{P^n_{\ca}\})$ as, otherwise, Lemma
\ref{l111} ensures that  $P_{\cc}\in \hat{L}(\{P^n_{\cc}\})$, a
contradiction (observe that, as $P\in\hat{\Pi}^{-1}(\hat{{\cal J}}(P_{\cc}))$ and $P^n\in\hat{\Pi}^{-1}(\hat{{\cal J}}(P^n_{\cc}))$, necessarily $P_{\ca}=I^-_{\ca}(P_{\cc})$ and
$P_{\ca}^n=I^-_{\ca}(P_{\cc}^n)$). So, Lemma \ref{auxiliar}
ensures the existence of a subsequence $\{P^{n_k}_{\ca}\}_{k}$ of
$\{P_{\ca}^{n}\}$ and $P_{\ca}\subsetneq P'_{\ca}\in
\hat{V}_{\ca}$ such that $P'_{\ca}\in \hat{L}(\{P_{\ca}^{n_k}\})$.
Now, taking into account that $P'_{\ca}=I^-_{\ca}(P'_{\cc})$ for
some $P'_{\cc}\in \hat{V}_{\cc}$ (apply Lemma \ref{lemmasobre}
with $g=g_{\ca}$) and applying Lemma \ref{l111}, we obtain that
$P'_{\cc}\in \hat{L}(\{P_{\cc}^{n_{k}}\})$ with $P'_{\cc}\neq
P_{\cc}$ (as $P'_{\ca}\neq P_{\ca}$). Finally, from the injectivity and continuity of $\apf$,
\[\apf(P_{\cc}),\apf(P'_{\cc})\in \LQ
(\{\apf(P^{n_k}_{\cc})\})\;\; \hbox{with}\;\;
\apf(P_{\cc})\neq\apf(P'_{\cc}),\] which contradicts the Hausdorff
character of $\hat{V}/\sim_{st}$. \cvd

\begin{remark}\label{rf1}{\em Under general circumstances (for example, open
subsets of manifolds) a continuous bijective map admits a
continuous inverse. However, the isocausality between $V$ and
$V_{\cc}$ does not imply the Hausdorffness of  $\hat V$ nor ${\hat V}/\sim_{st}$ and, obviously, the latter is an obstruction for $\hat {\cal J}$ being an
homeomorphism (recall that $\hat V_{\cc}$ is
Hausdorff). In order to
understand why $\hat V$ may be non-Hausdorff, recall that the
$\hat L$-convergence of a sequence of TIP's $\{P^n\}$ to some TIP
$P\subset \liminf \{P^n\}$ depends on the maximality of $P$ into
$\limsup \{P^n\}$. The hypotheses on $g_{\cc}$ (formulated by
using the Fermat metric for stationary spacetimes) prevent the
existence of different maximal sets in $\limsup \{P^n\}$, but
there is no reason to expect that such sets do not exist for $g$
-namely, a close but non-stationary metric. Nevertheless, the last
statement of Theorem \ref{apfc} shows that non-Hausdorffness is
the unique obstruction for the continuity of $\hat{\cal J}^{-1}$.
}
\end{remark}

\subsection{Past c-boundary $\check\partial V$}\label{j}
As we commented before, the corresponding arguments and results
for the past c-completion are deduced in a totally analogous way.
Summarizing Theorems \ref{l1} and \ref{apfc} we have:
\begin{theorem}\label{apfc'}
Let $V=(\R\times M,g)$ be a spacetime as in (\ref{e1}) such that
$g_{\cc}\prec_{0} g\prec_{0} g_{\ca}$ holds. If the integral
condition
%
\begin{equation}\label{e4'}\int_{-\infty}^0\left(\frac{1}{\replace{\sqrt{\alpha(s)}}{\alpha(s)}}-1\right)ds<\infty
\end{equation} holds, then the map
\[\begin{array}{cclc} \check{j}:& \check{V}_{\cc} & \rightarrow & \check{V} \\
& F_{\cc} & \mapsto & I^+(F_{\cc})\end{array}\] is injective. So,
$\check{V}$ contains a cone of base $M_{B}^{-}$ and apex $i^-$.

Moreover, if additionally $M_C^-$ is locally compact and $d_{Q}^-$ is a
generalized distance, then the map
\[
 \app =\check{\Pi}\circ\check{j}:\check{V}_{\cc} \rightarrow
\check{V}
/\sim_{st}
\]
is bijective and continuous. When $\check{V}/\sim_{st}$ is also
Hausdorff then $\app$ is an homeomorphism.
\end{theorem}

\begin{remark} {\em According to Corollary \ref{corollary1}, we can establish the
following assertions under the general hypotheses of Theorem
\ref{apfc'}: (1) if $(M,d^-)$ is bounded then $\check{\partial} V$
contains a cone of base $\partial_C^- M$ and apex $i^-$; if, in
addition, $(M,d^-)$ is complete (thus, compact), the past
c-boundary is $i^-$; (2) if $(M,d^-)$ is complete and (\ref{e4'})
holds, the past c-boundary contains a cone of base $\partial_{\cal
B}^- M$ and apex $i^-$ (recall Remark \ref{remarksection4}).
}
\end{remark}


\section{The (total) c-boundary $\partial V$} \label{s6}
Now, we are ready to study the total c-boundary of $V=(\R\times
M,g)$.
\subsection{$\partial V$ as a point set}
As in the study of partial boundaries (recall Theorems \ref{l1}
and \ref{apfc'}), firstly our aim is to introduce and analyze a
map of the form
\[\apt:\overline{V}_{\cc}\longrightarrow \overline{V}.\] In a first attempt, one
could try to define
$\apt((P_{\cc},F_{\cc}))=(\hat{j}(P_{\cc}),\check{j}(F_{\cc}))$.
However, it is not true in general that $P_{\cc}\sim_{S}F_{\cc}$
implies $\hat{j}(P_{\cc}) (=I^-(P_{\cc}))\sim_S
\check{j}(F_{\cc})(=I^+(F_{\cc}))$, and so, we cannot ensure that
$(P_{\cc},F_{\cc})\in \overline{V}_{\cc}$ implies
$(\hat{j}(P_{\cc}),\check{j}(F_{\cc}))\in \overline{V}$. The
reason of this problem is that, even if the choices for $\hat{j}$
and $\check{j}$ are canonical, there exists potential alternatives
(see Remark \ref{r1}), and the ambiguity derived from these
alternatives yields pathological properties when  the total
c-completion is studied. Therefore, in order to deal with $\apt$,
we have to look into the different alternatives for $\hat{j}$ and
$\check{j}$. With this aim, let us consider the following two
lemmas.

%

\begin{lemma}\label{p5} Let $(P_{\cc},F_{\cc})\in \overline{V}_{\cc}$ be such that $P_{\cc}$
and $F_{\cc}$ are associated to the same $p=(\Omega,x^s)\in
\R\times M_C^s$, i.e. $P_{\cc}=P_{\cc}(d^{+}_{p})$ and
$F_{\cc}=F_{\cc}(d^{-}_{p})$. Then, there exist $P\in \hat{V}$ and
$F\in \check{V}$ with $P\neq \emptyset\neq F$ such that $P\sim_S
F$, $P\in \hat{\Pi}^{-1}(\apf(P_{\cc}))$ and $F\in
\check{\Pi}^{-1}(\app(F_{\cc}))$. In the case that $x^s\in M$,
then $P=I^-((\Omega,x^s))$ and $F=I^+((\Omega,x^s))$.
\end{lemma}
{\it Proof.} The last assertion is trivial, so we can focus
directly on the case $x^s\in \partial_C^s M$, and thus,
$(P_{\cc},F_{\cc})\in \partial_{\cc} V$. As $P_{\cc}\sim_S
F_{\cc}$ and $g_{\cc}\prec g_{\ca}$, the following inclusions
hold:
\[ P_{\cc}\subset\downarrow_{\cc} F_{\cc} \subset\downarrow_{\ca}
F_{\cc},\;\; F_{\cc}\subset\uparrow_{\cc} P_{\cc}
\subset\uparrow_{\ca} P_{\cc}.\] Taking into account that
$g_{\cc}\prec_0 g$, we also have:
\[I^-(P_{\cc})\subset \downarrow I^+(F_{\cc})\qquad I^+(F_{\cc})\subset \uparrow I^-(P_{\cc}).\]
Therefore, since $I^{-}(P_{\cc})$ and $I^{+}(F_{\cc})$ are an IP
and an IF, resp., for $g$, there exist some TIP $P$ and some TIF
$F$ such that
\begin{equation}\label{eq5}I^-(P_{\cc})\subset P,\;\; I^+(F_{\cc})\subset F\;\; \hbox{ and }\;\; P\sim_{S}F\end{equation}
(for instance, take $P$ being maximal IP in $\downarrow
I^{+}(F_{\cc})$ containing $I^{-}(P_{\cc})$ and $F$ being maximal
IF in $\uparrow P$ containing $I^{+}(F_{\cc})$). Next, we are
going to prove that $P\in \hat{\Pi}^{-1}((\apf (P_{\cc}))$ and
$F\in \check{\Pi}^{-1}(\app (F_{\cc}))$. Taking into account that
$g\prec_0 g_{\ca}$, (\ref{eq5}) implies:
\[\begin{array}{c}I^-_{\ca}(P_{\cc})\subset I^-_{\ca}(P)\subset \downarrow I^+_{\ca}(F)\subset \downarrow I^+_{\ca}(F_{\cc})\\ I^+_{\ca}(F_{\cc})\subset I^+_{\ca}(F)\subset \uparrow I^-_{\ca}(P)\subset \uparrow I^-_{\ca}(P_{\cc}). \end{array}\]
In fact, for the first inclusion of the first line note that
$P_{\cc}\subset P$ implies $I^{-}_{\ca}(P_{\cc})\subset
I^{-}_{\ca}(P)$. For the second inclusion, $P\sim_{S}F$ implies
$P\subset \downarrow F$, and thus,
$I^{-}_{\ca}(P)\subset \downarrow_{\ca} I^{+}_{\ca}(F)$. Finally,
for the third inclusion, $F_{\cc}\subset F$ implies
$I^{+}_{\ca}(F_{\cc})\subset I^{+}_{\ca}(F)$, and so,
$\uparrow_{\ca}I^{+}_{\ca}(F)\subset
\uparrow_{\ca}I^{+}_{\ca}(F_{\cc})$. The inclusions in the second
line are deduced analogously.

From the construction, $I^-_{\ca}(P_{\cc})$ and
$I^+_{\ca}(F_{\cc})$ are associated to the same pair
$(\Omega,x^s)$, and so, $I^-_{\ca}(P_{\cc})\sim_S
I^+_{\ca}(F_{\cc})$.  From previous inclusions,
$I^-_{\ca}(P)=I^-_{\ca}(P_{\cc})$ and
$I^+_{\ca}(F)=I^+_{\ca}(F_{\cc})$, since, otherwise, the
maximality of $I^-_{\ca}(P_{\cc})$ and $I^+_{\ca}(F_{\cc})$ would
be violated (recall (\ref{eSz})). This joined to the fact that
$P_{\cc}\subset P$ and $F_{\cc}\subset F$ (recall (\ref{eq5}))
gives $P\in \hat{\Pi}^{-1}((\apf (P_{\cc}))$ and $F\in
\check{\Pi}^{-1}(\app (F_{\cc}))$, as required. \cvd

\begin{lemma}\label{ggg} Suppose that $d_Q^+$ is a generalized distance.
If $(P_{\cc},F_{\cc})\in \partial_{\cc} V$ satisfies
$F_{\cc}=\emptyset$ (resp. $P_{\cc}=\emptyset$), then
$(I^-(P_{\cc}),\emptyset)\in \partial V$ (resp. $(\emptyset,I^+(F_{\cc}))\in
\partial V$).

\end{lemma}
{\it Proof.} Suppose $F_{\cc}=\emptyset$ (thus, $P_{\cc}\neq
\emptyset$). As $d_Q^+$ is a generalized distance, there exists a
timelike curve $\gamma:[a,\infty)\rightarrow V$,
$\gamma(t)=(t,c(t))$, such that
$P_{\cc}=I^-_{\cc}(\gamma)=P_{\cc}(\bpcc{c})$ (recall the second
point of the Point Set structure in Theorem \ref{theo1}). In
particular, from $g\prec g_{\ca}$, \[\uparrow
I^-(P_{\cc})\subset\uparrow_{\ca} I^-_{\ca}(P_{\cc})=\emptyset.\]
Therefore, $(I^-(P_{\cc}),\emptyset)\in
\partial V$. \cvd

\smallskip

Observe that, under the hypothesis of $d_Q^+$ being a generalized
distance, previous lemmas include all the possibilities for pairs
in $\partial_{\cc} V$ (recall the second point of the Point Set
structure in Theorem \ref{theo1}). So, we are able to establish
the following result:

\begin{proposition} Let $V$ be a spacetime as in (\ref{e1}) such that
$g_{\cc}\prec_0 g\prec_0 g_{\ca}$, and assume that (\ref{e4}),
(\ref{e4'}) hold and $d_Q^+$ is a generalized distance. Consider
any map $\apt:\overline{V}_{\cc}\rightarrow \overline{V}$ given
by:

\begin{equation}\label{defapt}
\apt((P_{\cc},F_{\cc})):=\left\{\begin{array}{ll}
                          (I^-(P_{\cc}),\emptyset) & \hbox{ if
$F_{\cc}=\emptyset$}\\ (\emptyset,I^+(F_{\cc})) & \hbox{ if
$P_{\cc}=\emptyset$}\\ (P_0,F_0)   &
\hbox{otherwise}
                         \end{array}\right.
\end{equation}
where $(P_0, F_0)$ is any particular choice of pair obtained from
Lemma \ref{p5} (so that, $P_0\in \hat{\Pi}^{-1}(\apf(P_{\cc}))$
and $F_0\in \check{\Pi}^{-1}(\app(P_{\cc}))$). Then, $\apt$ is
well defined and injective.
\end{proposition}
{\it Proof. }
Observe that, if $d_Q^+$ is a generalized distance, any
$(P_{\cc},F_{\cc})\in \overline{V}_{\cc}$ should fall under the
conditions of Lemma \ref{p5} or Lemma \ref{ggg} (recall Theorem
\ref{theo1}, the second point about the Point Set structure), and
so, $\apt((P_{\cc},F_{\cc}))\in \overline{V}$, i.e., $\apt$ is
well defined.

For the injectivity, assume that
$\apt((P_{\cc},F_{\cc}))=(P,F)=\apt((P'_{\cc},F'_{\cc}))$. If
$P\neq\emptyset$ (the case $F\neq\emptyset$ is analogous), we
deduce from the construction that
\[P_{\ca}=I^-_{\ca}(P_{\cc})=I^-_{\ca}(P)=I^-_{\ca}(P'_{\cc})=P'_{\ca}\] So,
applying Proposition \ref{l11} (2) we have that $P_{\cc}=P'_{\cc}$. Finally, as
$d_Q^+$ is a generalized distance, Theorem \ref{theo1} ensures that
$F_{\cc}=F'_{\cc}$. \cvd

\smallskip

Recall that in the case of partial boundaries we made a canonical
choice for $\hat{j}$ and $\check{j}$, but alternative choices may
deserve to be taken into account (recall Remark \ref{r1}). Here,
$\apt$ is explicitly constructed by using any of the possible
$(P_0,F_0)$ provided by Lemma \ref{p5}.
In order to avoid this ambiguity, as well as possible problems
with the continuity of $\apt$, we will extend the concept of
strains to the c-completion (recall Definition \ref{strain}). In
this sense, let us associate first, to each $(P_{\cc},F_{\cc})\in
\overline{V}_{\cc}$, the subset
$\bstrain((P_{\cc},F_{\cc}))\subset \overline{V}$ defined as
follows:
\begin{equation}\label{bstrain}
(P,F)\in\bstrain((P_{\cc},F_{\cc}))\iff \left\{\begin{array}{lll}\hbox{if
$P\neq\emptyset\neq P_{\cc}$} & \hbox{then} & P\in
 \hat{\Pi}^{-1}(\apf(P_{\cc}))\\
 \hbox{if $F\neq\emptyset\neq F_{\cc}$} & \hbox{then} & F\in
 \check{\Pi}^{-1}(\app(F_{\cc}))\\ \hbox{if $P_{\cc}=\emptyset$} & \hbox{then} & P=\emptyset
\\ \hbox{if $F_{\cc}=\emptyset$} & \hbox{then} & F=\emptyset
 \end{array}
\right.
\end{equation}
In particular, if namely $(\emptyset,F)\in
\bstrain((\emptyset,F_{\cc}))$, then $F\neq \emptyset\neq
F_{\cc}$, and so, $F\in\check{\Pi}^{-1}(\app(F_{\cc}))$.
\begin{lemma}\label{ggg'}
If $(P_{\cc},F_{\cc})\neq (P'_{\cc},F'_{\cc})$ then
$\bstrain((P_{\cc},F_{\cc}))\cap \bstrain((P'_{\cc},F'_{\cc}))=\emptyset$.
\end{lemma}
{\it Proof.} Assume the existence of some
\[
(P,F)\in
\bstrain((P_{\cc},F_{\cc}))\cap\bstrain((P'_{\cc},F'_{\cc}))\neq\emptyset,\quad\hbox{with $P\neq \emptyset$,}
\]
(the case $F\neq \emptyset$ is
analogous). Then, $P\in \hat{\Pi}^{-1}(\apf(P_{\cc}))\cap
\hat{\Pi}^{-1}(\apf(P'_{\cc}))$ and, by the injectivity of $\hat{\Pi}$
and $\apf$, necessarily $P_{\cc}=P'_{\cc}$. On the other hand, as
$d_Q^+$ is a generalized distance, the c-completion $\overline{V}_{\cc}$ is simple as a point set
(recall Definition \ref{simpletop}). Therefore, from $P_{\cc}=P'_{\cc}$,
we deduce that $(P_{\cc},F_{\cc})=(P'_{\cc},F'_{\cc})$, as desired. \cvd

%

\smallskip

Now we are in conditions to introduce the following definition for
strains in $\overline{V}$ by means of a relation of equivalence
(observe that reflexivity and symmetry are straightforward, while
transitivity is a consequence of previous lemma).
\begin{definition}
Let $\relst$ be the relation of equivalence on
$\overline{V}$ defined by:
\[
 (P,F)\relst (P',F') \iff \left\{\begin{array}{l}(P,F)=(P',F'),\;
\hbox{ or}\\(P,F),(P',F')\in \bstrain
((P_{\cc},F_{\cc}))\quad\hbox{for some
$(P_{\cc},F_{\cc})\in\overline{V}_{\cc}$.}\end{array}\right.
\]
The (non-trivial) classes of equivalence in $\overline{V}/\relst$
will be called {\em (total) strains}. We will denote by
$\Pi:\overline{V}\rightarrow \overline{V}/\relst$ the natural
projection onto the quotient.
\end{definition}

\begin{figure}
\centering
\ifpdf
  \setlength{\unitlength}{1bp}%
  \begin{picture}(281.07, 185.65)(0,0)
  \put(0,0){\includegraphics{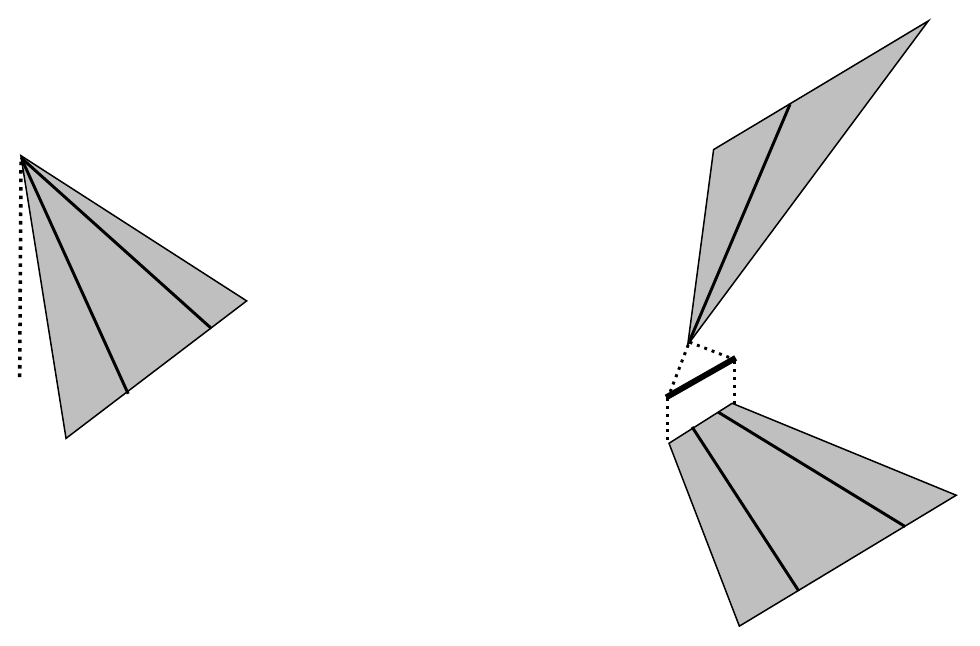}}
  \put(223.31,33.55){\fontsize{8.54}{10.24}\selectfont (A)}
  \put(225.20,139.73){\fontsize{8.54}{10.24}\selectfont (B)}
  \put(177.12,69.14){\fontsize{8.54}{10.24}\selectfont (C)}
  \end{picture}%
\else
  \setlength{\unitlength}{1bp}%
  \begin{picture}(281.07, 185.65)(0,0)
  \put(0,0){\includegraphics{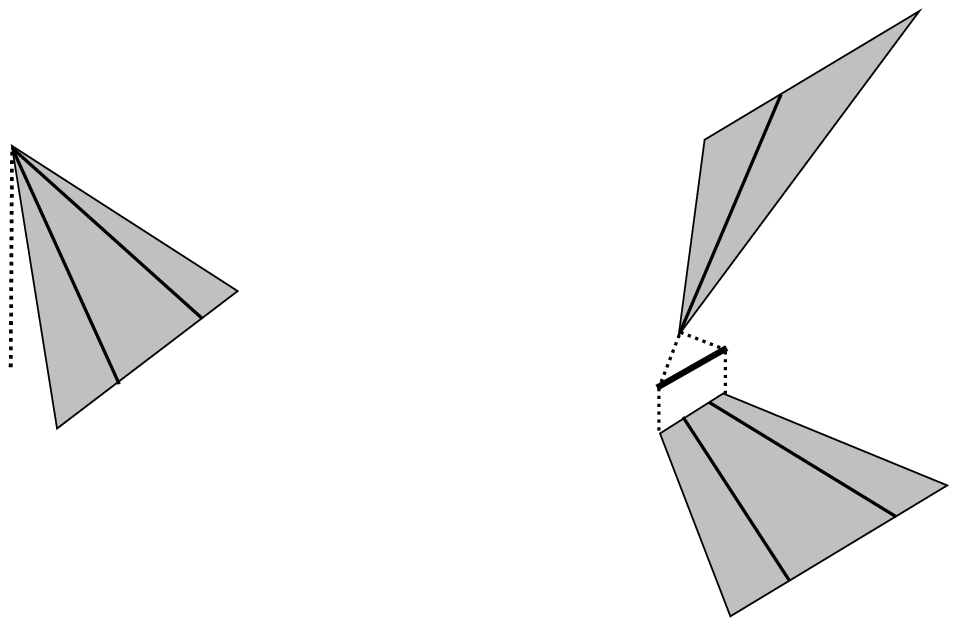}}
  \put(223.31,33.55){\fontsize{8.54}{10.24}\selectfont (A)}
  \put(225.20,139.73){\fontsize{8.54}{10.24}\selectfont (B)}
  \put(177.12,69.14){\fontsize{8.54}{10.24}\selectfont (C)}
  \end{picture}%
\fi
\caption{\label{figstraingeneral}{\it Two examples of (total) strains.}\newline The left figure shows a tame possibility for the strain. It is formed by a foliation of lightlike curves ending at the same point. All the boundary points in this strain are of the form $(P,\emptyset)$.\newline The right figure shows a wilder possibility. The strain is formed by three connected components: Two of them, labeled by (A) and (B), are foliated by lightlike curves, and the lightlike curves in the first one do not necessarily end at the same point. The third one is the segment (C). The boundary points in (A) and (B) are of the form $(P,\emptyset)$ and $(\emptyset,F)$ resp., while the points in the segment (C) are of the form $(P,F)$ with $P\neq \emptyset\neq F$.}
\end{figure}

%

\smallskip

As in the case of the partial boundaries, the following result
assures that the total boundary contains $\partial_{\cc}V$ as a
point set. Its proof is straightforward from Lemma \ref{ggg'}.
\begin{theorem}\label{th3}
Let $V=(\R\times M,g)$ be a spacetime as in (\ref{e1}) such that
$g_{\cc}\prec_{0} g\prec_{0} g_{\ca}$. If $d_Q^+$ is a generalized distance and
(\ref{e4}), (\ref{e4'}) hold, then the map
\begin{equation}
\label{ffff} \aptq: \overline{V}_{\cc}
\rightarrow\overline{V}/\relst, \quad (P_{\cc},F_{\cc})\mapsto
\Pi(\apt((P_{\cc},F_{\cc})))
\end{equation}
is injective.
\end{theorem}
%
%

Taking into account the structure of the c-boundary in terms of
lines (recall Theorem \ref{theo1}), we can also define {\em
quotient lines} in $\partial V/\relst$ of the following type:
\begin{equation}\label{lineaC}
 \aptq(\Lin(P_{\cc},F_{\cc})):=\{\aptq((P_{\cc}(\bpcc{c}+k),F_{\cc}(\bmcc{c'}
+k))) ,
k\in\R\},
\end{equation}
where $(P_{\cc},F_{\cc})\in \partial_{\cc}V$ with $P_{\cc}\equiv
P_{\cc}(\bpcc{c}),F_{\cc}=F_{\cc}(\bmcc{c'})$.

\noindent Such a definition suggests a natural lifting to
$\partial V$ defined by $\apt$ in the following way
\begin{equation}\label{columna}\apt(\Lin(P_{\cc},F_{\cc})):=\{\apt(P_{\cc}
(\bpcc{c}+k) , F_ { \cc } (\bmcc{c'}+k)) , k\in\R\}.
\end{equation}
This lifting preserves properties of the original line
$\Lin(P_{\cc},F_{\cc})$ with independence of the choice of $\apt$
in (\ref{defapt}). For example, as we will see in the next
section, it preserves the causal structure. Let us introduce some
notation for these elements:

\begin{definition}
Consider a line $\Lin(P_{\cc},F_{\cc})\subset \partial_{\cc}V$ for
some $(P_{\cc},F_{\cc})\in \partial_{\cc} V$ and consider the
quotient line $\aptq(\Lin(P_{\cc},F_{\cc}))\subset \partial
V/\relst$ (see (\ref{lineaC})). The set
$\apt(\Lin(P_{\cc},F_{\cc}))$, defined in (\ref{columna}), will be
called the {\em \columna} of the quotient line. For each $x\in
\aptq(\Lin(P_{\cc},F_{\cc}))$, $\Pi^{-1}(x)$ will be a {\em
\espina} of the quotient line. Finally, we will call the set
$\Pi^{-1}(\aptq(\Lin(P_{\cc},F_{\cc})))$ {\em \raspa} (see Figure \ref{fig3}).
\end{definition}
Finally, recalling the point set structure of $\partial_{\cc} V$
described in Theorem \ref{theo1}:
\begin{corollary}
Under the hypotheses of Theorem \ref{th3}, the boundary $\partial
V $ contains a {\raspa} (and thus, at least, a {\columna}) for
each line of $\partial_{\cc} V$.
\end{corollary}


\subsection{Causality}

For the causal structure, note that we have not already defined a
causal relation in the space $\overline{V}/\relst$. In fact, it is
not expected in general to obtain a complete satisfactory
definition of causal relation in this kind of spaces. The reason
is simple: we have not ensured that the relation between
representatives of different strains is independent of the choice
of such a representatives.
 However,
for $(P_{\cc},F_{\cc})\in \overline{V}$, the line
$\aptq(\Lin(P_{\cc},F_{\cc}))$ has a lifting, the {\columna},
which satisfies the following property.

\begin{theorem}\label{f}
Let $V=(\R\times M,g)$ be a spacetime as in (\ref{e1}) with
$g_{\cc}\prec_{0} g\prec_{0} g_{\ca}$. Assume that $d_Q^+$ is a
generalized distance and
(\ref{e4}), (\ref{e4'}) hold. For every
$(P_{\cc},F_{\cc})\in\partial_{\cc}V$, the \columna
\[
\apt(\Lin(P_{\cc},F_{\cc}))\;\;\hbox{is}\;\;\left\{\begin{array}{ll}
\hbox{horismotic} & \hbox{if, either $P_{\cc}=\emptyset$, or $F_{\cc}=\emptyset$} \\
\hbox{timelike} & \hbox{otherwise.}
\end{array}
\right.
\]
\end{theorem}
{\it Proof.} Assume that $P_{\cc}=P_{\cc}(\bpcc{c})$ and
$F_{\cc}=F_{\cc}(\bmcc{c'})$. Given two pairs
$(P_{\cc}^i,F_{\cc}^i)\in \Lin(P_{\cc},F_{\cc})$, $i=1,2$,
necessarily $P_{\cc}^i\equiv P_{\cc}(b_{c}^+ +k_i)$,
$F_{\cc}^i\equiv F_{\cc}(b_{c'}^- +k_i)$ with, say, $k_1<k_2$.
Then, one has two possibilities:
\begin{itemize}
\item Case $F_{\cc}=\emptyset\neq P_{\cc}$ (the case
$F_{\cc}\neq\emptyset=P_{\cc}$ is analogous). From Theorem
\ref{theo1} (with $\tilde{\alpha}\equiv 1$), $P_{\cc}^1\subset P_{\cc}^2$. Hence,
$I^-(P_{\cc}^1)\subset I^-(P_{\cc}^2)$, and thus
$(I^-(P_{\cc}^1),\emptyset)$ is causally, but not timelike,
related to $(I^-(P_{\cc}^2),\emptyset)$. In conclusion, the set
$\apt(\Lin(P_{\cc},F_{\cc}))$ is horismotic.

\item Case $P_{\cc}\neq \emptyset\neq F_{\cc}$. As
$\Lin(P_{\cc},F_{\cc})$ is timelike (see Theorem \ref{theo1}),
$P_{\cc}^2\cap F_{\cc}^1\neq \emptyset$. Consider
$(P^i,F^i)=\apt(P_{\cc}^i,F_{\cc}^i)$, with $i=1,2$.
From the
definition of $\apt$, $P^i_{\cc}\subset P^i, F^i_{\cc}\subset F^i$ (recall
Lemma \ref{p5}). Hence,
$\emptyset\neq F^1_{\cc}\cap P_{\cc}^2\subset F^1\cap P^2$, and
thus, $\apt(\Lin(P_{\cc},F_{\cc}))$ is timelike. \cvd
\end{itemize}
Even though there is not a unique choice for $\apt$, Theorem
\ref{f} is independent of this choice. So, the following
definition grasps the causal structure of the quotient
$\overline{V}/\relst$.
\begin{definition}\label{defcauqu}
The quotient line $\aptq(\Lin(P_{\cc},F_{\cc}))$ is: {\em horismotic} if
either
$P_{\cc}=\emptyset$, or $F_{\cc}=\emptyset$; and {\em timelike} otherwise.

\end{definition}

\subsection{Topology}
%

Now, we are ready to establish and prove our main result.
\begin{theorem}\label{theoremf}
Let $V=(\R\times M,g)$ be a spacetime as in (\ref{e1}) with
$g_{\cc}\prec_0 g\prec_0 g_{\ca}$. Assume that $d_Q^+$ is a
generalized distance, $M_C^s$ is locally compact and
(\ref{e4}),(\ref{e4'}) hold. Then, the map $\ap:\overline{V}_{\cc}\rightarrow
\overline{V}/\relst$ in (\ref{ffff}) is injective and
continuous.

If, in addition, $\overline{V}/\relst$ is Hausdorff, then $\ap$ is an
homeomorphism.

\end{theorem}

{\it Proof of the first statement.}
The injectivity of $\ap$ was proved in Theorem \ref{th3}. For the
continuity of $\ap$, assume that $(P_{\cc},F_{\cc})\in
L(\{(P_{\cc}^n,F_{\cc}^n)\})$ and let us prove
that $\{\ap
((P^n_{\cc},F^n_{\cc}))\}_{n}$ converges to $\ap
((P_{\cc},F_{\cc}))$.
By contradiction, suppose that there exists some neighborhood $U$ of
$\ap ((P_{\cc},F_{\cc}))$ such that, up to a subsequence, $\ap
((P^n_{\cc},F^n_{\cc}))\not\in U$ for any $n$. Then, the open set
$\Pi^{-1} (U)$, which contains all the elements in $\Pi^{-1}(\ap
((P_{\cc},F_{\cc})))$,  does not contain any element in
$\Pi^{-1}(\ap ((P_{\cc}^n,F_{\cc}^n)))$ for any $n$. Now, suppose
that $P_{\cc}\neq \emptyset$ (the case $F_{\cc}\neq \emptyset$ is
analogous), and thus, $P_{\cc}\in \hat{L}(\{P_{\cc}^n\})$. Note
that $P_{\cc}^n\neq \emptyset$ for $n$ big enough. From Lemma
\ref{propapf}, any sequence $\{P^n\}$, with $P^n\in
\hat{\Pi}^{-1}(\apf (P_{\cc}^n))$ for all $n$, admits some
subsequence $\{P^{n_k}\}_{k}$ converging to some $P\in
\hat{\Pi}^{-1}(\apf (P_{\cc}))$. Consider the following two
possibilities:
\begin{itemize}
\item Case $F_{\cc}=\emptyset$. Then, according to Lemma
\ref{ggg}, $(P,\emptyset)\in \Pi^{-1}(\ap ((P_{\cc},\emptyset)))$,
and so, $\{(P^{n_k},F^{n_k})\}_{k}\rightarrow (P,\emptyset)$ with
the chronological topology, where $F^{n_k}$ is chosen such that
$(P^{n_k},F^{n_k})\in \partial V$ for every $k$.

\item Case $F_{\cc}\neq \emptyset$. Then, $P^n_{\cc}\neq
\emptyset\neq F^{n}_{\cc}$ for $n$ big enough. From Lemma
\ref{p5}, one can take $(P^n,F^n)\in \Pi^{-1}(\ap
((P^n_{\cc},F^n_{\cc})))$ with $P^n\neq \emptyset\neq F^n$. From
Lemma \ref{propapf} (and its dual version for the past), there
exists some subsequence $\{(P^{n_k},F^{n_k})\}_{k}$ converging to
some $(P,F)\in \Pi^{-1}(\ap ((P_{\cc},F_{\cc})))$.
\end{itemize}
In both cases one deduces the existence of some subsequence
$\{(P^{n_k},F^{n_k})\}_{k}$, with $(P^{n_k},F^{n_k})\in
\Pi^{-1}(\ap ((P_{\cc}^{n_k},F_{\cc}^{n_k})))$, converging to some
$(P,F)\in \Pi^{-1}(\ap ((P_{\cc},F_{\cc})))$. So,
$(P^{n_k},F^{n_k})\in \Pi^{-1}(U)$ for $k$ big enough, a
contradiction. \cvd

\smallskip

In general, the result claimed in this first statement of previous
theorem is optimal. In fact, the structure of the c-completion of
the spacetime $(V,g)$ could be very complicated, and it may differ
radically from the c-completion of the stationary one (even after
the identification of the strains). So, in order to go further, it
is necessary to impose additional conditions which simplify this
structure. In this sense, the Hausdorff character of the quotient
space $\overline{V}/\relst$ imposed in the last statement of
Theorem \ref{theoremf} becomes natural (see
\cite{FH,FHSconf,FHSst} for other works where the Hausdorff
character of the c-completion has been analyzed).

In order to prove the last statement of Theorem \ref{theoremf},
first we establish the following technical lemma:
\begin{lemma}\label{lemmaauxiliar}
Assume that $d_Q^+$ is a generalized distance, $M_C^s$ is locally
compact and $\overline{V}/\relst$ is Hausdorff. If $(P,F)\in
\partial V$ with $P\neq\emptyset\neq F$, then
$P_{\ca}(=I^-_{\ca}(P))\sim_{S} F_{\ca}(=I^+_{\ca}(F))$.
\end{lemma}

{\it Proof.} Let $\gamma_P:[a,\Omega)\rightarrow V,
\gamma_P(t)=(t,c_P(t))$ and
$\gamma_F:[\overline{a},-\overline{\Omega})\rightarrow V,
\gamma_F(t)=(-t,c_F(t))$ timelike for $g$ such that
$P=I^-(\gamma_P)$ and $F=I^+(\gamma_F)$. From $\uparrow P\neq
\emptyset\neq \downarrow F$ (recall that $\emptyset\neq P\sim_S
F\neq\emptyset$), it is $\Omega, -\overline{\Omega}<\infty$.
Hence, $c_{P}$, $c_{F}$ must converge to some $x_0,x_1\in M_C^s$,
respectively (recall that $\gamma_P$, $\gamma_F$ are timelike
curves for $g_{\ca}$, the discussion above (\ref{ddd2}), Remark
\ref{remarksection4} and the fact that $d_Q^+$ is a generalized
distance). In particular,
$P_{\ca}=P_{\ca}(d^+_{(\Omega_{\ca},x_0)})$ and
$F_{\ca}=F_{\ca}(d^-_{(\overline{\Omega}_{\ca},x_1)})$, where
$\Omega_{\ca}$ and $\overline{\Omega}_{\ca}$ are defined as in
(\ref{omegaop}). In order to show that $P_{\ca}\sim_S F_{\ca}$, we
are going to prove that $(\Omega_{\ca},x_0)=(\overline{\Omega}_{\ca},x_1)$ (recall Remark \ref{remark1} assertion (i)), or, equivalently,
$(\Omega,x_0)=(\overline{\Omega},x_1)$.

Assume by contradiction that $\Omega\neq \overline{\Omega}$, and
thus, $\Omega<\overline{\Omega}$ (otherwise, $P\not\subset
\downarrow F$). From $P\sim_S F$, we ensure the existence of
sequences $\{t_n\}\subset [a,\Omega)$ and $\{s_n\}\subset
[\overline{a},-\overline{\Omega})$ with $t_n\nearrow \Omega,
s_n\nearrow -\overline{\Omega}$ and $\gamma_P(t_n)\ll
\gamma_F(s_n)$ for all $n\in \N$. Denote by
$\sigma_n:[t_n,s_n]\rightarrow V, \sigma_n(s)=(s,c_n(s))$ a
timelike curve joining $\gamma_P(t_n)$ and $\gamma_F(s_n)$.

First, note that for any sequence $\{r_n\}$ with $r_n\in
[t_n,s_n]$, it is $P\subset \LI(\{I^-(\sigma_n(r_n))\})$ and
$F\subset \LI(\{I^+(\sigma_n(r_n))\})$. Since $P\sim_S F$,
\cite[Lemma 3.15]{FHSconf} ensures that
\[
(P,F)\in L(\{(I^-(\sigma_n(r_n)),I^+(\sigma_n(r_n)))\}).
\]
In particular, from the definition of $\LQQ$ (recall (\ref{deflimitQ})),
\begin{equation}\label{sobreyectividadlema}
[(P,F)]\in \LQQ
(\{[(I^-(\sigma_n(r_n)),I^+(\sigma_n(r_n)))]\})\;\;\; \hbox{for
any $\{r_n\}$ with $r_n\in [t_n,s_n]$}.
\end{equation}
So, our aim will be to find a second limit point that contradicts
the Hausdorffness hypothesis. With this aim, we state the
following claim (proved below).

\smallskip

{\bf Claim:} {\em There exist $x_\infty\in M_C^s$, $t_\infty\in
(\Omega,\overline{\Omega})$ and a subsequence
$\{c_{n_k}\}\subset\{c_{n}\}$ such that $\lim_k
d_Q^+(c_{n_k}(t_\infty),x_\infty)=0$.}

\smallskip

In particular, since $d_Q^+$ is a generalized distance, we obtain
that $\{d^\pm_{(t_\infty,c_{n_k}(t_\infty))}\}$ converges
pointwise to $d^\pm_{(t_\infty,x_\infty)}$ and so that
$d^+_{(t_\infty,x_\infty)}\in
\hat{L}(\{d^+_{(t_\infty,c_{n_k}(t_\infty))}\})$ and
$d^-_{(t_\infty,x_\infty)}\in
\check{L}(\{d^-_{(t_\infty,c_{n_k}(t^\infty))}\})$ (recall
Proposition \ref{proposition1}). Therefore, if we consider the
terminal sets
$\overline{P_{\cc}}=P_{\cc}(d_{(t_\infty,x_\infty)}^+),
\overline{F_{\cc}}=F_{\cc}(d_{(t_\infty,x_\infty)}^-)$, the
characterization of $L$ at the beginning of Section \ref{s3.3} gives:
\[(\overline{P}_{\cc},\overline{F}_{\cc})\in
L(\{(I^-_{\cc}(\sigma_{n_k}(t_\infty)),I^+_{\cc}(\sigma_{n_k}(t_\infty))\}).
\]
Summarizing, if we apply the continuous function $\ap$ to previous
sequence, and taking into account (\ref{sobreyectividadlema}), we
deduce
\[
[(P,F)],\ap((\overline{P}_{\cc},\overline{F}_{\cc}))\in \LQQ
(\{(I^-(\sigma_{n_k}(t_\infty)),I^+(\sigma_{n_k}(t_\infty)))\}).
\]
So, in order to get the contradiction, it suffices to prove that $[(P,F)]\neq
\ap((\overline{P_{\cc}},\overline{F_{\cc}}))$ (recall that $\overline{V}/\relst$ is Hausdorff). Observe that, from the definition of $\relst$ and $\ap$, it is deduced directly that $\Pi^{-1}(\ap((\overline{P}_{\cc},\overline{F}_{\cc})))=\bstrain((\overline{P}_{\cc},\overline{F}_{\cc}))$. Then, one
only needs to prove that $(P,F)\notin \bstrain((\overline{P}_{\cc},\overline{F}_{\cc}))$ or,
particularly, that $P\notin \hat{\Pi}^{-1}(\apf(\overline{P}_{\cc}))$, since
$P\neq\emptyset\neq\overline{P}_{\cc}$. But this follows from the fact that $\overline{P}_{\cc}\not\subset P$. In
fact, $\overline{P}_{\cc}=P_{\cc}(d^{+}_{(t_{\infty},x_{\infty})})$ contains points $(t,x)\in V$ with $t>\Omega$
which cannot belong to $P=I^-(\gamma_P)$ (recall that the temporal component of $\gamma_{P}(t)$ is smaller than $\Omega$).

Therefore, we have obtained that $\Omega=\overline{\Omega}$.
Finally, as $P\subset \downarrow F$ and $g\prec_0 g_{\ca}$, we
have that $P_{\ca}\subset \downarrow_{\ca} F_{\ca}$. So, taking into account that $P_{\ca}=P_{\ca}(d^+_{(\Omega_{\ca},x_0)}), F_{\ca}=F_{\ca}(d^-_{(\Omega,x_1)})$ and (\ref{srelation}), we have that $P_{\ca}(d^+_{(\Omega_{\ca},x_0)})\subset P_{\ca}(d^+_{(\Omega_{\ca},x_1)})$, i.e.,
\[
d^+_{(\Omega_{\ca},x_0)}\leqslant d^+_{(\Omega_{\ca},x_1)}.
\] Evaluating both expressions in $x_1$, we deduce that
$d^+_Q(x_1,x_0)=0$ and, taking into account that $d_Q^+$ is a
generalized distance, necessarily $x_0=x_1$.

\smallskip

{\it Proof of the Claim. } Consider the auxiliary generalized
distance,
\[
\begin{array}{lccc}
d_{\aux}: & \left(\R\times M_C^s\right)\times \left(\R\times
M_C^s\right) & \rightarrow & \R
\end{array}
\]
with  \[d_{\aux}\left((s_1,y_1),(s_2,y_2)\right)=
d_Q^+(y_1,y_2)+|s_1-s_2|.\] Our aim is to apply \cite[Theorem
5.15]{FHSst} to the family of curves $\{\sigma_n\}$.  In order to
apply that theorem, we have to ensure the following conditions:
(i) the family of curves is defined in the same interval $I=[a,b]$
with $a<b$, (ii) $\sigma_n$ is an oriented equicontinuous family
of functions (see \cite[Definition 5.10]{FHSst}) and (iii)
$\sigma_n(a)$ has an accumulation point admitting a compact
neighborhood. For (i), observe that we can consider $a=\Omega$ and
$b=\overline{\Omega}$, as $\Omega<\overline{\Omega}, t_n\nearrow
\Omega$ and $s_n\searrow \overline{\Omega}$ .  For (ii), recall
that the curves $\sigma_n(s)=(s,c_n(s))$ are timelike for $g$, and
then, for $g_{\ca}$ (recall that $g\prec_0 g_{\ca}$). From
(\ref{charcau'}):
\[\replace{\sqrt{\alpha(s)}}{\alpha(s)}F^+(\dot{c}_n(s))<1,\] and so,
\begin{equation}\label{****}
 d^+(c_n(s_1),c_n(s_2))\leqslant \int_{s_1}^{s_2} F^+(\dot{c}_n(s))ds
<\int_{s_1}^{s_2} \frac{1}{\replace{\sqrt{\alpha(s)}}{\alpha(s)}}ds\leqslant
\frac{1}{{\rm min}|_{[t_0,s_0]}(\replace{\sqrt{\alpha(s)}}{\alpha(s)})}(s_2-s_1).
\end{equation}
Therefore, for $s_1<s_2$ we obtain that:
\[
 \begin{array}{rl} d_{\aux}(\sigma_n(s_1),\sigma_n(s_2)) & =(s_2-s_1) +
d^+(c_n(s_1),c_n(s_2))\\
 &
<(s_2-s_1)(1+\frac{1}{{\rm
min}|_{[t_0,s_0]}(\replace{\sqrt{\alpha(s)}}{\alpha(s)})}),\end{array}
\]
which imply that the family $\sigma_n$ is oriented equicontinuous.
Finally, for (iii), recall that
$d^+(c_n(t_n),c_n(\Omega))<\frac{1}{{\rm
min}|_{[t_0,s_0]}(\replace{\sqrt{\alpha(s)}}{\alpha(s)})}(\Omega-s_1)$
(see (\ref{****})), and so, as
$\sigma_n(t_n)=(t_n,c_n(t_n))\rightarrow (\Omega,x_0)$ and $d^+_Q$
is a generalized distance, it is deduced that
$\sigma_n(\Omega)=(\Omega,c_n(\Omega))\rightarrow (\Omega,x_0)\in
\R\times M_C^s$, being $M_C^s$ locally compact.

So, from \cite[Theorem 5.15]{FHSst} (recall that $M_C^s$ is
complete) there exist a subsequence $\{\sigma_{n_k}\}$ and a limit
curve $\sigma:[\Omega,\overline{\Omega}]$, such that $\lim_k
d_{\aux}(\sigma_{n_k}(s),\sigma(s))=0$ for all $s\in
[\Omega,\overline{\Omega}]$. Finally, the claim follows by taking
in previous limit $s$ as $t_\infty:=(\Omega+\overline{\Omega})/2$
and denoting $\sigma(t_\infty)=(t_\infty,x_\infty)$. \cvd

\smallskip

%
%
{\it Proof of the second statement.} It remains
to prove that
$\ap$ is surjective and has continuous inverse when
$\overline{V}/\relst$ is Hausdorff.

For the surjectivity of $\ap$, consider $[(P,F)]\in
\overline{V}/\relst$ and take an element $(P,F)\in \overline{V}$ of
the class $[(P,F)]$. From Lemma \ref{lemmasobre} (and its
corresponding past version), if $P\neq \emptyset$ (resp. $F\neq
\emptyset$) there exists a terminal set $P_{\cc}$ (resp.
$F_{\cc}$) for the metric $g_{\cc}$ such that $P_{\cc}\subset P$
with $P_{\ca}(:=I^-_{\ca}(P))=I^-_{\ca}(P_{\cc})$ (resp.
$F_{\cc}\subset F$ with $F_{\ca}(:=I^+_{\ca}(F))=I^+_{\ca}(F_{\cc})$). If $F=\emptyset$ (the
case $P=\emptyset$ is analogous), the stated properties of
$P_{\cc}$ ensure that $(P,\emptyset)\in
\bstrain((P_{\cc},F'_{\cc}))$ for any $F'_{\cc}$ such that
$\overline{V}_{\cc}$, and thus, $[(P,\emptyset)]=\ap
((P_{\cc},F'_{\cc}))$. If $P\neq \emptyset\neq F$, Lemma
\ref{lemmaauxiliar} ensures that $P_{\ca}\sim_S F_{\ca}$, and so,
$P_{\cc}\sim_S F_{\cc}$ (both sets are associated to the same
point in $\R\times M_C^s$, see Remark \ref{remark1} assertion (i)). In
particular, $(P,F)\in \bstrain((P_{\cc},F_{\cc}))$, and thus,
$[(P,F)]=\ap ((P_{\cc},F_{\cc}))$. In both cases,
$[(P,F)]=\ap((P_{\cc},F_{\cc}))$ for some $(P_{\cc},F_{\cc})\in
\overline{V}_{\cc}$.

%

For the continuity of $\ap^{-1}$, consider a sequence
$\{\ap((P_{\cc}^n,F_{\cc}^n))\}\subset \overline{V}/\relst$ and
some $\ap((P_{\cc},F_{\cc}))\in \overline{V}/\relst$ such that:
\begin{equation}\label{**1}\ap((P_{\cc},F_{\cc}))\in \LQQ
(\ap((P_{\cc}^n,F_{\cc}^n))).\end{equation}
By definition, there exist $(P^n,F^n)\in
\ap((P_{\cc}^n,F_{\cc}^n))$ and $(P,F)\in \ap((P_{\cc},F_{\cc}))$
such that:
\[(P,F)\in L(\{(P^n,F^n)\})
\]

We distinguish the following two cases:
\begin{itemize}
 \item Assume that $F=\emptyset$ (the case $P=\emptyset$ is analogous). From
$P\in \hat{L}(\{P^n\})$ and $g\prec_0 g_{\ca}$, we deduce that
$I^-_{\ca}(P)=P_{\ca}\subset \LI (\{P^n_{\ca}\})=\LI
(\{I^-_{\ca}(P^n)\})$. In order to prove that $P_{\ca}\in
\hat{L}(\{P_{\ca}^{n}\})$, assume by contradiction that $P_{\ca}$ is not maximal into $\LS (\{P^n_{\ca}\})$.
From Lemma \ref{auxiliar} there exists some $P'_{\ca}$ such that
$P_{\ca}\varsubsetneq P'_{\ca}\in \hat{L}(\{P_{\ca}^{n_k}\})$ for some
subsequence $\{P_{\ca}^{n_k}\}\subset \{P_{\ca}^n\}$.
From Lemma \ref{l111}, we obtain that $P'_{\cc}\in
\hat{L}(\{P_{\cc}^{n_k}\})$, where $P'_{\cc}$ satisfies $I^-_{\ca}(P'_{\cc})=P'_{\ca}$ (recall Lemma \ref{lemmasobre} with $g=g_{\ca}$). From the last assertion in Theorem
\ref{theo1}, $(P'_{\cc},F'_{\cc})\in
L(\{(P_{\cc}^{n_k},F_{\cc}^{n_k})\})$ for any $F'_{\cc}\in
\check{V}$ such that $(P'_{\cc},F'_{\cc})\in \overline{V}$. The
continuity of $\ap$ ensures that
\begin{equation}\label{**2}\ap((P'_{\cc},F'_{\cc}))\in \LQQ
(\{\ap((P_{\cc}^{n_k},F_{\cc}^{n_k}))\}).\end{equation} Therefore, from
(\ref{**1}), (\ref{**2}) and the Hausdorffness of $\overline{V}/\relst$, we
have
that
$\ap((P_{\cc},F_{\cc}))=\ap( (P'_{\cc},F'_{\cc}))$. So,
$P'_{\cc}=P_{\cc}$, and thus, $P'_{\ca}=P_{\ca}$ a contradiction.

So, $P_{\ca}\in \hat{L}(\{P^n_{\ca}\})$ and from Lemma \ref{l111}, $P_{\cc}\in \hat{L}(\{P^n_{\cc}\})$. Finally, from the last
assertion of Theorem \ref{theo1}, $(P_{\cc},F_{\cc})\in
L(\{(P_{\cc}^n,F_{\cc}^n)\})$.

 \item Assume that $P\neq\emptyset\neq F$. From $g\prec_0 g_{\ca}$, we deduce that
$P_{\ca}\subset \LI (\{P_{\ca}^n\})$ and $F_{\ca}\in \LI(\{F_{\ca}^n\})$.
Lemma \ref{lemmaauxiliar} ensures that $P_{\ca}\sim_S F_{\ca}$, and so,
$(P_{\ca},F_{\ca})\in L(\{(P_{\ca}^n,F_{\ca}^n)\})$ (recall
\cite[Proposition 3.16]{FHSconf}). Hence, from Lemma \ref{l111},
\[(P_{\cc},F_{\cc})\in L(\{(P_{\cc}^n,F_{\cc}^n)\}).\] \cvd
\end{itemize}

\begin{remark}\label{rf2}{\em In contraposition to the maps $\hat
{\cal J}, \check{\cal J}$ for the partial boundaries, now the map
$ {\cal J}$ might be non-surjective. This is a consequence of the
fact that a TIP $P$ may be S-related to multiple TIF's $F, F'$
(and viceversa), yielding more than one point in the quotient.
However, these points must be non-Hausdorff related, and so, the
Hausdorffness of $\overline{V}$ (or just of $\overline{V}/\relst
$) would be sufficient to ensure surjectivity. At any case, if one
redefine the relation of equivalence $\relst$ by redefining
previously the strains in (\ref{bstrain}) as follows,
\begin{equation}\label{bstrainP}
(P,F)\in\bstrainp((P_{\cc},F_{\cc}))\iff
\left\{\begin{array}{ll}
P\in\hat{\Pi}^{-1}(\apf(P_{\cc})) & \hbox{if $P\neq \emptyset$}\\
 F\in \check{\Pi}^{-1}(\app(F_{\cc})) & \hbox{if $P=\emptyset$}
\end{array}\right.,
\end{equation}
the new projection ${\cal J}$ will be bijective (with no
assumption on Hausdorffness). Note however that this definition
introduces an asymmetry with respect to the temporal orientation,
as it is focused on the $P$ part. In fact, we can consider a dual
definition $\bstrainm$ based analogously on the $F$ part,
obtaining a relation of equivalence which differs from
(\ref{bstrainP}) (and (\ref{bstrain})).

Finally, we remark that, independently of the approach considered,
i.e. the original one for $\bstrain$ or the time asymmetric ones
$\bstrainp$ and $\bstrainm$, the assertions in Remark \ref{rf1}
about the possible lack of continuity of ${\cal J}^{-1}$ in the
non-Hausdorff case remain true. }\end{remark}

\section{Conclusions}

By using simple standard conformally stationary spacetimes (i.e.,
spacetimes as in (\ref{e1}) under the hypotheses of Theorem
\ref{theoremf})
as model spaces, we have proved that isocausality yields the
qualitative behavior of the c-boundary for a wide class of
spacetimes. Such spacetimes include those in the general split
form $V=\R\times M$ (see (\ref{e1}), and recall that conformally
related spacetimes can be also taken into account), under the
relevant hypothesis that the metric will stabilize for large
values of $|t|$, in the sense of (\ref{e4}) and (\ref{e4'}).

Our approach has been developed in full generality by using
stationary spacetimes rather than static ones. This is remarkable,
as one expects that the evolution of most isolated systems will
stabilize them into stationary spacetimes. The approach is
applicable as the c-boundary of standard stationary spacetimes is
known \cite{FHSst} (including the stationary part of Kerr
spacetime \cite{FlHerr}).

Such a broad viewpoint leads to consider some technical properties
of the generalized distances $d^\pm$ for the model spacetime and
their extensions $d_Q^\pm$ to the Cauchy boundaries. But these
properties will hold trivially in practical cases. For example, if
we consider as model spacetime a static one $V_{\cc}=(\R\times M,
g_{\cc}=\Lambda (-dt^2+h))$ then  no hypothesis on $d_Q^\pm$ is
necessary, and we have just to check that the Cauchy boundary for
$d^+$ (a usual distance equal to $d^-$ in the static case) is
locally compact. Obviously, this is a natural property for any
model spacetime
---and will always hold in a realistic spacetime.

The results obtained in this paper can be extended to other cases.
For example,
 we have assumed for the model stationary spacetime the structure
$\R\times M$ (i.e., the timelike Killing vector field $\partial_t$
is complete), but one can consider also the case $I\times M$ where
$I\subset \R$ is an interval. This is interesting, as all the
causal elements are conformally invariant, and  spacetimes such as
 Robertson-Walker ones lie in this class (up to a conformal
transformation). This computation would be somewhat long, but it
can be carried out with the introduced tools and, consequently, it
has not been done here.

Summing up, the techniques introduced here complete the study of
the c-boundary in \cite{FHSconf, FHSst}, by making its qualitative
behavior understandable even in situations with scarce symmetries.

\section*{Acknowledgements}

The authors are partially supported by the Spanish MICINN Grant
MTM2010-18099 and Regional J. Andaluc\'ia Grant P09-FQM-4496, both
with FEDER funds.


\begin{thebibliography}{99}

%

\bibitem{BEE} J.K. Beem, P.E. Ehrlich, K.L. Easley,
{\em Global Lorentzian geometry}, Monographs Textbooks Pure Appl.
Math. {\bf 202} (Dekker Inc., New York, 1996).


%

%
%


\bibitem{CJS} E. Caponio, M.A. Javaloyes, M. S\'anchez, On the
interplay between Lorentzian Causality and Finsler metrics of
Randers type, {\em Rev. Matem. Iberoamericana} {\bf 27} (2011)
919--952.




\bibitem{FH} J.L. Flores, S.G. Harris, Topology of the
causal boundary for standard static spacetimes, {\em Class. Quant.
Grav.} {\bf 24} (2007), no. 5, 1211--1260.

\bibitem{FlHerr} J.L. Flores, J. Herrera, The c-boundary construction of
spacetimes: application to stationary Kerr spacetime, {\it VI Int.
Meeting on Lorentzian Geometry}, Granada, 6-9 September, 2011.

\bibitem{FHSnota}J.L. Flores, J. Herrera, M. S\'anchez, Isocausal spacetimes may
have different causal boundaries, {\em Class. Quant. Grav.} {\bf
28} 175016.

\bibitem{FHSconf}  J.L. Flores, J. Herrera, M. S\'anchez, On the final definition of the causal boundary and its relation with the conformal
boundary, {\em Adv. Theor. Math. Phys.} {\bf 15} (2011) 991-1058.

\bibitem{FHSst}  J.L. Flores, J. Herrera, M. S\'anchez, Gromov, Cauchy and causal boundaries for Riemannian, Finslerian and
Lorentzian manifolds (2010), to appear in {\em Memoirs of the
AMS}. Available at arXiv:1011.1154.

\bibitem{FS} J.L. Flores, M. S\'anchez,
 The causal boundary of wave-type spacetimes, {\em J. High Energy Phys.} (2008), no. 3, 036, 43 pp


\bibitem{GP-Sa}
A. Garc{\'\i}a-Parrado, M. S\'anchez, Further properties of causal
relationship: causal structure stability, new criteria for
isocausality and counterexamples, {\it  Class. Quant. Grav.} {\bf
22} (2005) 4589--4619.


\bibitem{GpScqg03} A. Garc\'{\i}a-Parrado, J. M. M. Senovilla, Causal
relationship: a new tool for the causal characterization
of Lorentzian manifolds, {\em Class. Quant. Grav.} {\bf 22} (2003)
625--664.

\bibitem{GSa} A. Garc\'{i}a-Parrado, J.M.M. Senovilla, Causal
symmetries, {\em Class. Quant. Grav.} {\bf 20} (2003) L139.

\bibitem{GSb} A. Garc\'{i}a-Parrado, J.M.M. Senovilla, General study and basic
properties of causal symmetries, {\em Class. Quant. Grav.} {\bf 21} (2004)
661--696.

\bibitem{GpScqg05}
A. Garc{\'\i}a-Parrado, J. M. M. Senovilla, Causal structures and
causal  boundaries, {\it Class. Quant. Grav.} {\bf 22} (2005)
R1--R84.


\bibitem{GKP}
R.P. Geroch, E.H. Kronheimer and R. Penrose, Ideal points in
spacetime, {\it Proc. Roy. Soc. Lond. A} {\bf 237} (1972) 545--67.

%
%
%
%
%
%


%
%
%
%
%

\bibitem{MS} E. Minguzzi, M. S\'{a}nchez,  The causal hierarchy of
spacetimes, in {\em Recent developments in pseudo-Riemannian Geometry} (2008) 359--418. 
ESI Lect. in Math. Phys., European Mathematical Society Publishing
House. (Available at gr-qc/0609119).

%

\bibitem{O} B. O'Neill, {\em Semi-Riemannian Geometry with applications to Relativity}, Academic Press, INC, 1983.

%
%
%
%
%
%
%
%

\bibitem{Sz}
L.B. Szabados, Causal boundary for strongly causal spaces, {\it
Class. Quant. Grav.} {\bf 5} (1988) 121--34.

%




\end{thebibliography}
\end{document}